\documentclass[10pt]{article}

\usepackage[colorinlistoftodos,prependcaption,textsize=tiny]{todonotes}
\usepackage{multirow}
\usepackage{multicol}
\usepackage{amsmath}
\usepackage{amssymb}
\usepackage{float}
\usepackage[hyphens]{url}
\usepackage{hyperref}

\usepackage{color}
\usepackage{graphicx}
\definecolor{dkgreen}{rgb}{0,0.6,0}
\definecolor{gray}{rgb}{0.5,0.5,0.5}
\definecolor{mauve}{rgb}{0.58,0,0.82}
\definecolor{truepositive}{HTML}{FE0000}
\definecolor{falsepositive}{HTML}{DAA520}
\newcommand{\K}{$\mathbb{K}$}

\newtheorem{definition}{Definition}

\usepackage{url}
\usepackage{listings}

\lstdefinestyle{JavaStyle} {language=Java}
\begin{document}

\title{Detection of App Collusion Potential Using Logic Programming}

%\author{\IEEEauthorblockN{Jorge~Blasco,~\IEEEmembership{Member,~IEEE,} Thomas~M.~Chen,~\IEEEmembership{Senior member,~IEEE,} Igor~Muttik,~\IEEEmembership{Member,~IEEE,} and ~Markus~Roggenbach}
\author{Jorge~Blasco, Thomas~M.~Chen, Igor~Muttik and \\
~Markus~Roggenbach}

%\IEEEauthorblockN{\\
%\IEEEauthorrefmark{1}Jorge.Blasco.1@city.ac.uk,
%\IEEEauthorrefmark{2}Tom.Chen.1@city.ac.uk,
%\IEEEauthorrefmark{3}igor.muttik@intel.com,
%\IEEEauthorrefmark{4}m.roggenbach@swansea.ac.uk}
%}
%\markboth{IEEE Transactions on Mobile Computing,~Vol.~XX, No.~Y, Month~Year}%
%{Blasco \MakeLowercase{\textit{et al.}}: Efficient Detection of Android App Collusion Potential Using Logic Programming}
%\markboth{IEEE Transactions on Mobile Computing,~Vol.~XX, No.~Y, Month~Yar}%

%\IEEEtitleabstractindextext{%

%\begin{IEEEkeywords}
%D.4.6 Security and Privacy Protection, J.9 Mobile Applications, C.2.8.b Architectures, K.6.5.c Invasive software
%\end{IEEEkeywords}}

\maketitle

\begin{abstract}
%TODO: (Markus) Review the abstract being more explicit about the originality, significance and rigour of the paper.
Android is designed with a number of built-in security features such as app sandboxing and permission-based access controls. Android supports multiple communication methods for apps to cooperate. This creates a security risk of app collusion. For instance, a sandboxed app with permission to access sensitive data might leak that data to another sandboxed app with access to the internet. 
In this paper, we present a method to detect potential collusion between apps. 
First, we extract from apps all information about their accesses to protected resources and communications. Then we identify sets of apps that might be colluding by using rules in first order logic codified in Prolog. After these, more computationally demanding approaches like taint analysis can focus on the identified sets that show collusion potential. This "filtering" approach is validated against a dataset of manually crafted colluding apps. We also demonstrate that our tool scales by running it on a set of more than 50,000 apps collected in the wild. Our tool allowed us to detect a large set of real apps that used collusion as a synchronization method to maximize the effects of a payload that was injected into all of them via the same SDK.
\end{abstract}
%\IEEEdisplaynontitleabstractindextext

%\IEEEpeerreviewmaketitle

% READY TO REVIEW
%****************************************
%     INTRODUCTION
%****************************************
%\IEEEraisesectionheading{
\section{Introduction}\label{sec:introduction}
%}

% Motivation of Collusion problem. It is important bla bla...
% 
%\IEEEPARstart {S}{martphones}
Smartphones are pervasive in modern everyday life. The number of smartphones in use is predicted to grow from 2.6 billion in 2016 to 6.1 billion in 2020 \cite{lunden2015}. One reason for this fast adoption is the extensive ecosystem of apps for taking photos, sending messages, banking, and millions of other familiar uses. Smartphones hold a great deal of personal information (e.g., photos, finances, credentials, messages) making them very appealing targets for criminals who commonly use malicious apps to steal sensitive information \cite{lipovsky2014eset}, extort users \cite{kovacs2015}, or misuse the device services for their own purposes \cite{page2015}.

To mitigate these threats, mobile operating systems offer a multi-sandbox environment where each app is executed in isolation from the rest. This isolation is intended to limit the effects of any potential malicious activity of an app. A way to get around this restriction is to execute a privilege escalation exploit or persuade the user to grant additional permissions to the app which usually works quite well because users are generally unaware of the risks associated with granting permissions to apps \cite{felt2012android}. %With the improvements of the permission system in Android, single app malware will have more trouble to dissuade the user into accepting an extensive list of permissions\footnote{Since Android 6.0 permissions are granted separately at runtime when required, not altogether at once during installation}. 
%To mitigate this, Google now offers a service that scan apps uploaded to their market for security vulnerabilities and threats. Additionally, all phones with the Google version of Android, include a service that scans the apps behavior in the background to detect malicious activity. However, as malware protections get better, malware is also getting smarter, more complex, and more difficult to detect  \cite{sophos205}. 

% Unfortunately, even when a security savvy user is in charge, sandboxed operating systems can still be attacked without the need of developing expensive privilege escalation exploits.
These restrictions can also be circumvented by app collusion. 
%However,  security model is focused on protecting against threats model only consider threats malicious applications can take advantage of the Android permission model to affect devices owned by security savvy users. 
In collusion, the malicious activity is split across multiple apps coordinated through inter-app communications (IAC). 
This kind of attack is possible because sandboxed systems, such as Android, are designed to prevent threats from individual apps. However, they do not restrict or monitor inter-app communications and therefore they fail to protect from multiple apps cooperating in order to achieve a malicious goal. Most malware analysis systems, such as those that power antivirus software for smartphones, also check apps individually. While collusion is not common today, it opens an avenue to circumvent the security of Android and other sandboxed operating systems. And the risk of attackers taking this "collusion avenue" is likely to grow because of constant enhancements in efficacy of static and dynamic analysis of individual apps as well as in calculating their reputations.

% permission restrictions that could be easily exploited by criminals to become a serious threat in the near future. 

%TODO (Jorge)- Add a subsection within the introduction that talks about the contributions of the paper. Be more explicit about this. In the related work section, compare related works with our contributions.
\subsection{Contributions}
%TODO (Tom)- Clarify the scope of our paper and mention that we cannot use other approaches such as anomaly detection because of the nature of the problem itself. There is no ground truth to compare against and we cannot cover everything. Do this in the contribution section.
In this paper, we present a methodology to detect the potential for app collusion by using logic programming. For each app, we extract what we call an ASR (Access-Send-Receive) signature. Then Prolog rules are used to characterize app collusion behaviours. These rules reflect two aspects of collusion: access to protected resources and communication channels between apps. Using the rules and the set of ASR signatures codified as Prolog facts, it is possible to obtain a list of potentially colluding app sets from large datasets of apps. We have validated the approach against a set of manually crafted colluding apps and then applied it to another set of more than 50,000 apps collected in the wild. Our approach has allowed us to shed light on how apps in the Android ecosystem communicate and it allowed us to identify a set of apps in the wild all having an embedded malicious SDK that used collusion to maximize the effects of a malicious payload. As with any rule-based method, our approach has limitations to known types of collusion. In principle, an approach such as anomaly detection might be able to detect new types of collusion that known rules can not, but it is currently not possible to test or compare anomaly detection until new collusion attacks occur (or become known). %These applications were found to be on the wild  We have identified a set of applications insiUsing our approach we have been able to identify unkown colluding behaviour This methodology has allowed us to identify Using this methodology we have found that a famous malicious SDK, known as MoPlus SDK, uses collusion to execute its functionality in the app with the higher privileges in the device.
%Overapproximation because we do not pretend to prove that it happens, but flag the suspicious sets so they can be further analysed to check if it really happens or not. 
%As another contribution of this paper, we also provide a detailed analysis of how real applications use some of their communication channels. This information could be useful to derive other collusion detection methods in the field.

The remainder of this paper is structured as follows. Section \ref{sec:android} reviews the Android security model focusing mainly on the aspects related to app collusion. 
% This section can be of special interest for readers not familiarized with the Android platform. 
In Section \ref{sec:collusion} we propose a definition of app collusion and describe possible communication channels that could be used by colluding apps. %Section \ref{sec:improving} describes our arquitecture to incorporate multiple applications into current malware detection systems, which are prepared only to deal with single application malware. 
Section \ref{sec:b-m} describes our approach in detail and gives  
validation of our approach against a set of manually created colluding apps. Section \ref{sec:s-u} elaborates on how our method can scale up. Section \ref{sec:experiments} offers experimental results with a large dataset of apps in the wild. Section \ref{sec:moplus} reports a group of colluding apps found in the wild, that were flagged by our method. To the best of our knowledge, this is the first example of app collusion found in the wild. Section \ref{sec:related} reviews previous work done to detect and protect against app collusion. 
% Finally, section \ref{sec:conclusion} presents a summary and the conclusions obtained after analysing the results obtained in this paper.
% comment: don't really need to say last section is Conclusions. 

%****************************************
%     Android Security Model
%****************************************
\section{The Android Operating System}
\label{sec:android}
% 1-2 page/s. 

%This section briefly describes the main components of Android applications, its communications and the foundations of its security model. 

The Android operating system model is designed to protect users, apps, the device and the network from malicious parties. By default, any third party app is considered untrusted by the OS and runs inside a sandbox that isolates it from any sensitive resource or other apps. Until Android 4.3, the sandboxing mechanism was implemented by assigning a different Linux user id (UID) to each app and configuring file permissions accordingly. Since Android 4.4 SELinux domains are used in addition to the Linux UID so apps can only access files inside its sandbox, as these are the only ones in its domain.  
Access to sensitive resources outside the app sandbox is possible by using APIs provided by the operating system. Calls to these APIs are managed by a permission system, which has a deny-by-default policy. Apps that want to access sensitive resources, must include a permission declaration inside its \lstinline|AndroidManifest.xml| file. When the app is being installed (in Android versions below 6.0), the system will ask the user to accept the permissions used by the app before proceeding with the installation. At this point, the user must accept or deny all permissions requested by the app. Starting with Android 6.0, apps can ask for permissions at runtime and users have the choice of granting or denying each permission. However, permissions must still be declared in the manifest file. For a more detailed description of Android security features, we refer the reader to \cite{enck2009understanding,shabtai2009securing}.

\subsection{App Components}

Android apps are built with the following components:
\begin{itemize}
\item \textit{Activities} represent screens of the user interface and allow the user to interact with the app. Activities run only in the foreground. Apps are generally composed of a set of activities.
\item \textit{Services} execute operations in the background. They are generally used by other components of the app to perform long-running tasks: listening to incoming connections, downloading a file, etc. %Services can be configured so they can be accessed remotely by other processes.
\item \textit{Broadcast receivers} respond to messages that are sent through \lstinline{Intent} objects, by the same or other apps. 
\item \textit{Content providers} manage access of other apps to the app's own data. Apps with content providers enable other apps to read and write their local data.  
\end{itemize}

In order to be reachable by other apps, components must be declared as \emph{exported} in the app manifest file. %This file includes all the necessary information that the OS requires to execute and handle the different application components. Additionally, developers can program some parts of their apps using native C code that is compiled to each of the specific platform architectures available in Android. 

\subsection{Communications }

%Applications in Android can use standard Unix communication mechanisms like files, sockets, etc. However, these are heavily restricted by SELinux since Android 4.4. As an example, even if an application sets its own file permissions for other users to read and write files in its sandbox, no other application will actually be able to access them, as they are being confined to another SELinux domain. 

Besides the standard Unix files, sockets, etc., Android offers three inter-process communication (IPC) mechanisms: 
\begin{itemize}
\item The \textit{Binder} is a remote procedure call mechanism designed to enable fast and efficient IPC between processes that run inside a sandbox. The Binder Framework uses a server-client architecture. It is implemented as a Linux driver, allowing communications between sandbox boundaries.% by  Apps offer services that can be consumed by other apps through the Binder. The communication between apps is mediated by the Binder Driver, which is a component of the Binder framework that is implemented as a Linux driver in the Android kernel. 
This allows the operating system to mediate communication across sandbox boundaries. The rest of the Android IPC (Intents and Content Providers) are, in fact, abstractions based on the Binder.% In fact, even the Android component lifecycle management calls are executed via the Binder.
\item An \textit{Intent} is a messaging object which is used to request actions from other apps' components. These can belong to the same or different apps. Intents can be explicit or implicit. Explicit intents target specific activities or services. Implicit intents target generic actions that can be performed by many different activities (send a message, open a web link, etc.). Activities, services and broadcast receivers declare the intents which they can handle by declaring a set of \lstinline|IntentFilters|. For activities and services, intent filters must be declared in the app's manifest. Broadcast receivers can also register their intent filters programmatically during execution.
\item \textit{Content Providers} are used to offer other apps a method to access their own structured data. Content providers store information in one or more tables, similarly to relational databases. Apps access data of content providers using \lstinline|ContentResolver| objects. A content provider offers methods, which can be called by other apps, not only to read data, but also to update, create and delete information encapsulated in the content provider object.
\end{itemize}

None of these three IPC mechanisms, which all are standard in Android, is covered by the mandatory access control offered by SELinux \cite{mutti2015selinux}. As a result, apps can share any kind of information by using standard IPC without any restriction. To avoid security problems, Android allows apps communicating through IPC to request specific permissions to any app that wants to communicate with them. As an example, Android includes a Contact Provider to interact with the device's contact list. Apps accessing this provider need to declare \lstinline|READ_CONTACTS| or \lstinline|WRITE_CONTACTS| permissions in their manifest. Similarly, apps using Intents to start phone calls require the  \lstinline|CALL_PHONE| permission. %The goal of this mechanism is to avoid exposing access to sensitive resources through public interfaces.

Unfortunately, Android does not enforce this protection mechanism. It is left to app developers to decide if they want to apply it. Consequently, permission-protected resources are potentially exposed. This fact can be exploited to build colluding apps which access sensitive resources without permission. Apps might also communicate in order to aggregate permissions necessary to perform malicious actions. %In addition to this, malware analysis services as well as security researchers normally focus on single apps. Therefore, they cannot detect when malicious actions are distributed over several apps installed on a device. 

%****************************************
%     Application Collusion
%****************************************
\section{App Collusion}
\label{sec:collusion}
%TODO A6(Jorge)- Remove mentions to false positives and state the there is no ground truth. We are analysing an unknown problem (with real world apps). Our effort is not to detect and compare against a ground truth, but to extract knowledge from a set of apps we are analysing that might or might not be colluding. Say that there is no state of the art that defines a ground truth of colluding apps we can test againts.

The first mention of app collusion was \emph{Soundcomber}, a proof-of-concept malware described in 2011 \cite{schlegel2011soundcomber}. It was comprised of two apps which used inter-app communications to steal the user's banking credentials. Soundcomber shows the limitations of the Android permission model to protect against apps that collude to aggregate their permissions \cite{bugiel2011xmandroid}. Although collusion has inherently a malicious component, sometimes it is hard to distinguish collusion from collaboration.

A frontal attack on detecting collusion of pairs, triplets, and larger sets of Android apps is not practical given the search space. Thus, in this work, we aimed to develop an effective filtering system to quickly analyse large app sets to detect collusion potential so a security analyst or other more computationally expensive automatic tools can focus their efforts on the most suspicious app sets. 

%Hence, we do not try to strictly detect a collusion attack. 

%The malicious component of collusion is also acknowledged in a recent set of papers from Bagheri, Sadeghi et al. \cite{bagheri2015covert1,sadeghi2015analysis}. A more detailed definition of collusion is given by  Kalim O. Elish \cite{elish2015user,elish2015need}. He defines a collusion attack as the collaboration between malicious applications, likely written by the same adversary, to obtain a set of permissions to perform malicious attacks. Kalim differentiates two types of collusion, according to the goal of the adversary: Collusion for data leakage and collusion for system abuse.  

%Literature on app collusion can be traced back to the \emph{confused deputy} attack \cite{hardy1988confused}. Confused deputies expose protected resources through public interfaces. In Android, confused deputy attack may take a form of \emph{permission re-delegation attack} \cite{felt2011permission, davi2011privilege, wu2015effective}. A careless developer may unintentionally expose permission-protected resources by allowing a component to access those resources by communicating with other apps through IPC. 
% This is advantageous for the attacker because the malicious app does not have to declare the usage of the protected resources.
%Soundcomber

\subsection{Definition of Collusion Potential }
%\subsection{Collusion Definition}
%A careful definition of collusion is needed to account for subtle differences between collusion, confused deputies, and app cooperation. 
% To the best of our knowledge there is no reference in the previous literature about collusion attacks that have been observed in the wild. Nevertheless, malicious behaviours similar to collusion are evident from known cases of apps exploiting insecure exposure of sensitive data by other apps \cite{arzt2014flowdroid}. Differentiating collusion attacks from confused deputies and permission re-delegation attacks can be very difficult. Although the motivation and original purpose of the apps involved in the attacks are different, they may result in the same effects. 
In this work, collusion refers to the ability of a set of apps to carry out an attack through collaboration. 
%This is in contrast to most existing literature where collusion is usually associated with inter-app communications for permission re-delegation. 
We assume that colluding apps can carry out the same malicious actions as single apps such as information theft, money theft, service misuse or sabotage \cite{suarez2013evolution}.
%\begin{itemize}
%\item Information theft: when one app accesses sensitive information and another app sends that information outside the device.
%\item Money theft: when two apps collaborate to abuse a device cost-sensitive APIs like placing calls or sending SMS messages.
%\item Service misuse: when one app is able to control a system service but receives commands from another app to control it.
%\end{itemize}

To the best of the authors' knowledge there is no evidence that collusion can create new threats in the mobile scenario, as they depend on the assets, which remain constant. However, collusion can change how attacks are executed. In the Soundcomber scenario, collusion is used to make an information leakage attack more stealthy. Additionally, malicious apps can also take advantage of collusion just for coordination and synchronization. %In this case, each app can be considered as malicious on its own. When the colluding apps are installed in the same device, they coordinate their actions to enhance their impact (compared to acting on their own). %Using the methodology proposed in this paper, we have detected apps in the wild that implemented this kind of collusion (see Section \ref{sec:moplus}). We believe this behaviour passed unnoticed because researchers were not giving attention to this more subtle kind of collusion.
Considering this, we define collusion potential based on the following:
\begin{itemize} 

\item[A1:] Actions are operations provided by the operating system (Android) API. % such as record audio, send data through the Internet, receive data from another app, etc. %Let $Act$ denote the set of all actions.
%%%%% or set of all permissible actions ? 
We consider three kinds of actions. \textbf{A}ccess actions involve access to system-protected resources (e.g. record audio). \textbf{S}end actions allow apps to send information to other apps on the same device. \textbf{R}eceive actions allow apps to receive information from other apps. Actions are grouped in what we call the Access-Send-Receive (ASR) signature of an app, denoted by $ASR_{app}$.% is a triplet $(A_{app},S_{app},R_{app})$ where $A_{app}$ represents the access of $app$ to system-protected resources, and $S_{app}$ and $R_{app}$ are the capabilities of $app$ to send and receive information (using inter-app communications), respectively. %through any available Android communication channel and $R_{app}$ represents the capabilities of $app$ to receive information from other app using any available Android communication channel. Note that we consider a capability as the ability to execute an action.

\item[A2:] Actions can be characterized by a number of attributes such as permissions, API calls, etc. In this work we use static analysis to extract attributes. %Let $B$ denote the set of all action attributes and $pms:Act \to \wp(B)$ specify the set of permissions required by Android to execute an action.

\item[A3:] A threat $t$ can be described by a sequence of actions $t=\langle a_1,a_2, \cdots, a_n\rangle$. In this work, we consider the threats to be the ones that can be created by single apps. Let $\tau$ denote the set of all these threats.

%\item[A4:] An inter-app communication $c$ is a set of actions. Let $com$ denote the set of all known inter-app communications.
%%%%%% set of all permissible inter-app communications ?

\end{itemize}

\begin{definition}[Collusion Potential]\label{def:collusion}
A set $S$ consisting of at least two apps has collusion potential if the apps in $S$ together can execute a sequence $seq$ of actions such that:
\begin{enumerate}

\item $seq$ restricted to \textit{Access} actions is a sequence in $\tau$; furthermore, $seq$ is collectively executed involving all apps in $S$, i.e., each app in $S$ executes at least one action in $seq$; and

\item considering the \textit{Send} and \textit{Receive} actions in $seq$, all apps in $S$ are connected through communication channels. That is, it is possible to build a directed graph $G=(S, C)$ where the elements of $C$ are pairs of \textit{Send} and \textit{Receive} actions, in which there are no unreachable nodes, i.e.\ apps in $S$.

\end{enumerate} 

\end{definition}

Our definition of collusion potential highlights two steps required for collusion: execution of a malicious action (threat) and the need to communicate between the apps executing these actions. Malicious actions and types of threats that can be executed by smartphone malware have been extensively studied by researchers \cite{suarez2013evolution,la2013survey}. 
Due to the high level description of the actions it may happen that app sets marked as having collusion potential are taking advantage of other apps (permission re-delegation attacks \cite{felt2011permission}) or just collaborating. In this work we consider that both cases should be detected and highlighted as having collusion potential. As already noted by other researchers \cite{fang2014permission}, the difference between a malicious or non-malicious behaviours can only be seen by comparing the application descriptions, developer intentions  (which are difficult to measure by static analysis) and system implementations. So, in all these cases, apps can exhibit the exact same behaviour, and therefore must be considered to have collusion potential (i.e. in all three cases apps have the capability to collude). It is up to the security analysis or the taint analysis tools executed when app sets show collusion potential to decided whether one app is executing a permission re-delegation attack over another app, they are collaboration with the user knowledge or they are actually colluding. 

In the next section we review the main communication channels that can be used by colluding apps. The development of the definition into a Prolog program to detect collusion potential is described in section \ref{sec:b-m}.
                
%$S_{app}$ represents the sources of information that generates and application and $R_{app}$ represents the sinks of the app, as described by Marforio et al. \cite{marforio2011application}. However, it must be noted that we do not use their notation in the exactly same sense. Marforio et al. only consider sources and sinks as channels that are only used for transmission of sensitive information. In our approach, these channels can be used to transmit anything, from sensitive information, to a command that executes a specific task in another application. 

\subsection{Communication Channels for App Collusion}\label{ssec:communication}

Colluding apps require some form of communication to execute and/or synchronize their actions. Colluding apps can use standard communication channels (as described previously in Section \ref{sec:android}) or stealthy communications (often also called "covert communications") to avoid being detected. The following communication mechanisms may be employed:

\begin{itemize}
\item \textit{Intents} can be used by colluding apps to share information. Broadcast receivers and services allow apps to exchange data without user intervention. %In addition, intents used to open activities can include information that is not necessarily required to present the new activity to the user. 
\item Malicious apps can use \textit{content providers} as a dropbox to exchange information. %This runs the risk of being visible to the user (e.g. creating a new contact to exchange information). 
Access to system content providers requires apps to request permissions (e.g. \lstinline|WRITE_CONTACTS| for the contact database). 
\item \textit{External storage} of an Android device can also be used as a shared dropbox to exchange information. 
%External storage is generally available through a USB connection, SD card or even via non-removable storage. 
Apps accessing the external storage need to declare the \lstinline|READ_EXTERNAL_STORAGE| or \lstinline|WRITE_EXTERNAL_STORAGE|, depending on the required access. % can write to and read from external storage. 
Files in the external storage can be accessed using the common file access API. 
\item \textit{Shared preferences} are an Android feature that allows apps to store key-value pairs of data. %Its purpose is to store app configuration and preferences. 
Although it is not intended for inter-app communication, apps can use key-value pairs to exchange information if proper flags are defined (\lstinline|WORLD_READABLE| or \lstinline|WORLD_WRITABLE|) when accessing and storing data. Since the adoption of SELinux apps cannot access the world readable files of other apps, as they are confined to different SELinux domains. 
\item Colluding apps can also use standard \textit{Unix sockets} to communicate through the local network interface. % Apps can use sockets opened to localhost to communicate as if they were communicating through the network. 
Communication between two apps that is mediated by an external server is not generally counted as collusion, because the communication happens outside the device domain. 
\item \textit{Covert channels} may take advantage of APIs or features offered by the operating system to enable communication between processes \cite{schlegel2011soundcomber,marforio2012analysis}. In Android, this includes publicly readable and writable settings (e.g. volume level) and capturing broadcast intents generated by the system. Additionally, processes can take advantage of covert channels present in most computing systems like file locks, process enumeration, free space and CPU usage.

\end{itemize}
 
% \subsection{App Collusion in the Literature}

%In this work we focus on the usage of overt channels to enable collusion

%TODO Review if this should be included
%****************************************
%     Improving app analysis process for collusion detection
%****************************************
%\section{Improving app analysis process for collusion detection}
%\label{sec:improving}
% Half a page. 
% Present old model. Describe problems
% Add diagrams.
% Present new model. How this improves. Some problems of this model maybe
% 
% Storing - for buying & warning

%****************************************
%     Methodology
%****************************************
\section{Detecting Collusion Potential}
\label{sec:b-m}
%TODO (Jorge)- Describe more explicitly what is exactly automatic in our paper. The filtering is automatic. Further analysis can be automatic (model checking) or manual, as in the case of the MoPlus SDK.

%TODO A6(Jorge)- Remove mentions to false positives and state the there is no ground truth. We are analysing an unknown problem (with real world apps). Our effort is not to detect and compare against a ground truth, but to extract knowledge from a set of apps we are analysing that might or might not be colluding. Say that there is no state of the art that defines a ground truth of colluding apps we can test againts.

%TODO (Joge) Strengthen (not much) the fact that our approach is a filter, not a detector. Provide hints about other ways of continuing to work with potentially colluding pairs (Model checking).

Mobile apps can be downloaded from the web or app markets such as ``Google Play.'' 
These apps are analysed by market operators and anti-malware services that constantly crawl these markets. A combination of static and dynamic analysis techniques as well as statistical methods are used to establish reputation and risk values for an app. However, all these techniques consider apps in isolation and neglect to take into account other apps that could be installed on the same device. This hinders the detection of possible collusion behaviours from an appropriate combination of apps installed on the same device. 

Our approach aims to extend app analysis services by also considering the collusion potential of sets of apps; the reputation and risk of an app are measured not only in terms of its own features, but also factoring in its capabilities when installed with other apps on the same device. 

%Our tool development address two challenges: in Section \ref{sec:b-m} on "Basic Methodology" we demonstrate how to automatically find sets of potentially colluding apps; in Section \ref{sec:s-u} on "Scaling up" we address the question of how to deal with large sets of apps.

%Our methodology is based on a scenario where a user wants to check if the apps installed on his device have any collusion potential.
To do so, for each app we extract the actions it is able to perform as ASR signatures. These are described as Prolog facts. In a similar way, collusion potential is described as a set of logic rules that are composed by Prolog facts. %Once all app capabilities have been extracted, 
Collusion rules can be applied to query for apps that may be potentially colluding. A detailed view of the process is given in the following sections. 

%It (1) uses Androguard~\cite{desnos2013androguard} to extract facts about
%the communication channels and permissions of all single apps in a
%given app set $S$, (2) which is then abstracted into an
%over-approximation of actions and communication channels that could be used by a single app. (3) Finally the collusion rules are fired if the proper combinations of actions and communications are found in $S$.

\subsection{Extracting ASR Signatures}

We extract ASR signatures by performing a static analysis of the app manifest and app code. In this work we consider the usage of (i) implicit \emph{intents}, (ii) \emph{shared preferences} and (iii) \emph{external storage} for communication, i.e., a subset of the channels listed in Section \ref{ssec:communication}. Therefore, the ASR signature of an app is a combination of all permissions, intents, shared preferences and external storage channels that can be used by an app to send or receive information. %It must be noted that, as described in Section \ref{sec:collusion}, an app can create several communication channels using each of these mechanisms. For instance, an app that stores information in two shared preferences files, would be creating two communication channels.%, one for each file. 

To analyse app code, we have extended Androguard \cite{desnos2013androguard}, a reverse engineering tool for Android apps written in Python\footnote{Our code is available at \url{https://github.com/acidrepo/collusion_potential_detector}}. Our extension looks up API calls involved in the creation and broadcast of intents and broadcast receivers and the access and modification of shared preferences files. Parameters that specify the communication channel for each method are tracked back through the code. We trace back the value of the \lstinline|action| parameter for each broadcast intent and corresponding intent filters. In the case of shared preferences, we track the name of the shared preferences file. As with any static analysis tool, our tool is not able to trace back values that are dynamically defined. In those cases, we return the API call path that generates the value. 

The app manifest is analysed to identify usage of external storage or static broadcast receivers. We are not able to obtain the specific channel used by apps through external storage. This requires identifying all API calls that can modify the external storage file system. PScout is able to identify the Android library API calls that require \lstinline|READ_EXTERNAL_STORAGE| and \lstinline|WRITE_EXTERNAL_STORAGE| permissions \cite{au2012pscout}. However, standard IO calls are not included in this mapping. Therefore we have left this task for future work. 

As an example, Table \ref{tab:sig_appsms} shows the ASR signature of a simple artificial colluding app that would send premium SMS messages. This app requires \lstinline|READ_SMS| to display the user's SMS and \lstinline|SEND_SMS| to send SMS. This permission is abused to send SMS messages when an intent with the action \lstinline|sms| is received. %This app has been developed along with other apps to test our methodology. 
Details about this app, and the rest of the apps created for validation purposes, can be found on section \ref{sec:testappset}.
%Example of a communication signature (mathematical description, table, and maybe with an image). One app 
\begin{table}[t]
\centering
\caption{ASR signature of the SMS app (id 4) that is part of the Botnet group.}
\label{tab:sig_appsms}
\begin{tabular}{|l|l|}
\hline
\multicolumn{2}{|c|}{$A_{id4}$} \\
\hline
\multirow{2}{*}{Permissions} & \lstinline|READ_SMS|\\
\cline{2-2}
& \lstinline|SEND_SMS| \\
\hline
\hline
\multicolumn{2}{|c|}{$S_{id4} = \emptyset$} \\
\hline
\hline
\multicolumn{2}{|c|}{$R_{id4}$} \\
\hline
\multirow{2}{*}{Intent} & \lstinline|SMS_SENT| \\
\cline{2-2}
 & \lstinline|sms| \\
%\hline
%Shared Preferences & -\\
%\hline
%External Storage & -\\
\hline
\end{tabular}
\end{table}

%Table \ref{tab:sig_appsharer} shows the ASR signature of the non-colluding sharer app (id 11). In this case, the app uses is able to capture sharing broadcast intents with the \lstinline|android.intent.action.SEND| action. The app requires access to the internet to share the data included in the intent with an internet service. This behavior is similar than the one offered by multiple messaging apps like Whatsapp, Telegram, Facebook Messenger, etc.

%\begin{table}[t]
%\centering
%\caption{ASR signature of the non-colluding sharing app (id 11).}
%\label{tab:sig_appsharer}
%\begin{tabular}{|l|l|}
%\hline
%\multicolumn{2}{|c|}{$A_{id11}$} \\
%\hline
%Permissions & \lstinline|INTERNET| \\
%\hline
%\hline
%\multicolumn{2}{|c|}{$S_{id11} = \emptyset$} \\
%\hline
%\hline
%\multicolumn{2}{|c|}{$R_{id11}$} \\
%\hline
%Intent & \lstinline|android.intent.action.SEND| \\
%\hline
%Shared Preferences & -\\
%\hline
%External Storage & -\\
%\hline
%\end{tabular}
%\end{table}

\subsection{Characterizing Collusion Potential with Logic Rules}

Our approach to detect collusion potential utilises logic programming in Prolog. We have created a Prolog program, the \emph{ACiD} (\textbf{A}pplication \textbf{C}ollus\textbf{i}on \textbf{D}etection) rule set, that defines when a set of applications shows collusion potential.%, depending on their ASR signatures.% The ASR signatures extracted from applications are translated into Prolog facts to feed the ACiD rule set. Its goal is to serve as a fast and computationally cheap filter to detect potentially colluding apps. 

%Our action set $Act$ is composed of three kinds of actions: \textit{access}, \textit{send} or \textit{receive}. 
\textit{Access} actions have been categorized into four high level actions: accessing sensitive information; use an API that can incur a financial loss; control device services (e.g. camera); and send information outside the device. These actions are characterised by permissions and API calls which are mapped to one or more of the four high level actions. For example, an app that declares the \lstinline|INTERNET| permission will be capable of sending information outside the device:
\begin{equation*}
uses(App,P_{Internet}) \rightarrow information\_outside(App)
%\label{eq:internetpermission}
\end{equation*}
Similarly:
\begin{eqnarray*}
uses(App,P_{Read\ contacts}) &\rightarrow& sensitive\_info(App) \\
uses(App,P_{Send\ SMS}) &\rightarrow& money(App) \\
uses(App,P_{Kill\ process}) &\rightarrow& control\_service(App)
\end{eqnarray*}

Overall, in Android 4.3 35 permissions can be used to access sensitive information; 12 to send information outside the device; 3 to execute financially-sensitive APIs; and 39 to control device services. The complete mapping of permissions to actions can be found on the project repository.%\footnote{\url{https://github.com/acidrepo/collusion_potential_detector/blob/master/collusion_rules.pl}}.

\textit{Send} and \textit{Receive} actions are characterised by specific API calls offered by the Android OS. For each we create a fact that describes that communication action. When using \lstinline|Intents| and \lstinline|SharedPreferences| we are able to specify the communication channel using the intent actions and preference file respectively. As an example, if an app sends a
\lstinline|BroadcastIntent| with an action \lstinline|SEND_FILE| we
consider the following:
\begin{eqnarray*}
send\_broadcast(App,Intent_{send\_file}) \\
\rightarrow send(App,Intent_{send\_file})
\end{eqnarray*}
We consider that two apps communicate if one of them is able to $send$ and the other to $receive$ via the same channel. 
\begin{eqnarray*}
send(App_a,channel) \wedge receive(App_b,channel) \rightarrow   \\  
 communicate(App_a,App_b,channel)
\end{eqnarray*}
Note that communication is directed, i.e., information flows from $App_a$ to $App_b$.

%\subsection{Threats}
Finally, each of the threats is characterised by a sequence of actions. Our threat set $\tau$ considers information theft, money theft and service misuse. Specifically, we consider that two apps have collusion potential to execute an information theft when one of them has access to sensitive information and communicates with another app which can do external communications:
\begin{eqnarray*}
            sensitive\_info(App_a) &\wedge& \\
            information\_outside(App_b) & \wedge &  \\
            communicate(App_a,App_b,channel) & \rightarrow & \\ 
 			information\_collusion(App_a,App_b) &&
\end{eqnarray*}

We consider that two apps have potential to collude for money theft when one app has access to cost sensitive APIs and receives information from another app:
\begin{eqnarray*}
            money(App_b) & \wedge &  \\
            communicate(App_a,App_b,channel) & \rightarrow & \\ 
 			money\_collusion(App_a,App_b) &&
\end{eqnarray*}

An internet connection in $App_a$ would allow a server to send commands to an app with access to cost sensitive APIs:
\begin{eqnarray*}
            information\_outside(App_a) &\wedge& \\
            money(App_b) & \wedge &  \\
            communicate(App_a,App_b,channel) & \rightarrow & \\ 
 			money\_collusion(App_a,App_b) &&
\end{eqnarray*}

In a similar sense, this same app could also send commands from a C\&C server to other apps that have access to device services:
\begin{eqnarray*}
            information\_outside(App_a) &\wedge& \\
            control\_service(App_b) & \wedge &  \\
            communicate(App_a,App_b,channel) & \rightarrow & \\ 
 			service\_collusion(App_a,App_b) &&
\end{eqnarray*}

\subsection{ACiD Rule Set in Prolog}

We have translated the ACiD rules into a Prolog program. These include the rules required to identify communication paths (and specific channels) between applications. Then, once the ASR signatures have been extracted from an app set, they can be translated into Prolog facts to be part of the Prolog program that is executed to find collusion potential.

For every permission used by an app a Prolog fact \lstinline|uses(app)| is created. For every channel sending information outside the app sandbox we generate a \lstinline|send(app,channel)| fact and \lstinline|recv(app,channel)| for all channels receiving information from outside the sandbox.

A Prolog predicate (\lstinline|q :- p|) describes a logical rule of the form $p \rightarrow q$. Prolog uses \emph{modus ponens} to evaluate queries and look for results. If \lstinline|p| is true, then it will consider \lstinline|q| to be also true. 
The identification of communication paths between apps is performed by using recursive Prolog predicates (Listing \ref{lst:comm}). The base case (first rule) identifies when two apps are communicating. This is, if \lstinline|AppA| sends information through \lstinline|Channel| and \lstinline|AppB| receives information from the same channel, it means they communicate (\lstinline|comm_l(AppA,AppB,2,_,[]|). The recursive predicates (last two rules) add more apps to the communication path. To avoid circular paths, all rules check if the app that is being analysed is already a member of the path (\lstinline|nonmember|).% These rules allow to query for paths of specific length and an initial app.

\begin{lstlisting}[caption={Communication rules},label={lst:comm},basicstyle=\scriptsize]
comm_l(AppA,AppB,2,_,[]) :- trans(AppA,Channel), recv(AppB,Channel), AppA\=AppB.				
comm_l(AppA,AppB,Length,[],[AppD|Rest]) :- Length > 2,trans(AppA,Channel),	recv(AppD,Channel), AppA\=AppD, PrevL is Length -1, comm_l(AppD,AppB,PrevL,[AppA],Rest), AppA\=AppB.			
comm_l(AppA,AppB,Length,Visited,[AppD|Rest]):- Length > 2, trans(AppA,Channel), recv(AppD,Channel), AppA\=AppD, nonmember(AppD,Visited), PrevL is Length - 1, comm_l(AppD,AppB,PrevL,[AppA|Visited],Rest), AppA\=AppB.
\end{lstlisting}

The channel rules allow, once a collusion path has been obtained, to extract the list of communication channels used by the apps (Listing \ref{lst:chan}). Similarly, the first predicate (first rule) saves the channel used when two apps are communicating. The second predicate looks recursively for the rest of the channels
%if there are more than two apps colluding. 
These rules facilitate the security analyst task to investigate how the potential collusion can happen.

\begin{lstlisting}[caption={Channel identification rules},label={lst:chan},basicstyle=\scriptsize]
chanl(AppA,AppB,[],Channel) :- trans(AppA,Channel),recv(AppB,Channel), AppA\=AppB.
chanl(AppA,AppB,[AppD|Rest],[Channel|Channels]) :- trans(AppA,Channel), recv(AppD,Channel), chanl(AppD,AppB,Rest,Channels). 
\end{lstlisting}

\subsection{Validation}
\label{sec:testappset}
% 10 apps + collusion results
% Describe stage one as finding collusion patterns between small sets of apps. 

We have performed an initial validation by running it through a set of eleven specifically developed artificial apps that include colluding and non-colluding apps\footnote{Due to the malicious nature of the apps, they are only available upon request.} which are summarized in Table \ref{tab:collusion_summary}. We have decided to use apps developed by us for two reasons. First, to the best of our knowledge, no colluding apps have been identified in the wild yet. Thus, we lack a set of previously known positive examples. Second, even if there were apps identified as colluding, we could not be 100\% certain on non-collusion: even an app downloaded from a reputable market might be colluding. i.e., we lack a set of previously known negative examples.

There are nine colluding apps that have been developed to cover all collusion scenarios described in Section \ref{sec:collusion}. They can be categorized in three groups:
\begin{itemize}
\item The \textbf{Document Extractor} group is composed of two apps. One of the apps in the group looks for sensitive documents (txt, pdf, db, xls, etc.) on the external storage ($app_1$). This information is shared with $app_2$ using the shared preferences. The information received is sent to a remote server.
\item The \textbf{Botnet} group is composed of four apps. One of the apps ($app_4$) acts as a relay that receives orders from the command and control center. The other colluding apps execute commands depending on their requested permissions. They are capable of sending SMS messages ($app_4$), stealing the user's contacts ($app_5$) and starting and stopping tasks ($app_6$). This group uses intents as communication channel.
\item The \textbf{Contact Extractor} group is composed of three apps. This group sends the device's address book to a remote server. The first app ($app_7$) reads the contacts from the address book, the second ($app_8$) forwards them to the third ($app_9$), which sends them to the Internet. This group uses intents and the external storage as communication channels.
\end{itemize}

In addition to the colluding apps, the validation set includes two non-colluding apps. These are a document viewer ($app_{10}$) and an information sharing app ($app_{11}$). The first app displays different file types on the device screen and uses other apps (through an intent with the action \lstinline|android.intent.action.SEND|) to share their uniform resource identifier. The second app receives text (through the same action) and sends it to a remote server.

\begin{table*}[t]
\centering
\caption{Summary of colluding app groups included in our basic app set.}
\label{tab:collusion_summary}
\begin{tabular}{|l|l|l|l|l|l|}
\hline
\textbf{Group} & \textbf{Id} & \textbf{Threats} & \textbf{Permissions} & \textbf{Colludes with} & \textbf{Channel} \\
\hline
Document& 1 &  \multirow{2}{*}{Information Theft} & \lstinline|READ_EXTERNAL_STORAGE|& 2 & Shared Prefs. \\
\cline{2-2}\cline{4-6}
Extractor & 2 &  & \lstinline|INTERNET| & 1 & Shared Prefs. \\
\hline
\hline
\multirow{6}{*}{Botnet} & 3 & \multirow{2}{*}{Information theft} & \lstinline|INTERNET| & 4,5,6 & Intents\\
\cline{2-2}\cline{4-6}
& \multirow{2}{*}{4} &  &  \lstinline|READ_SMS|& \multirow{2}{*}{3} & \multirow{2}{*}{Intents} \\
&  & \multirow{2}{*}{Service misuse} &  \lstinline|SEND_SMS|& & \\
\cline{2-2}\cline{4-6}
& 5 & & \lstinline|READ_CONTACTS| & 3 & Intents \\
\cline{2-2}\cline{4-6}
& \multirow{2}{*}{6} & \multirow{2}{*}{Money theft}  & \lstinline|GET_TASKS|&\multirow{2}{*}{3} & \multirow{2}{*}{Intents}\\
&  &  & \lstinline|KILL_BACKGROUND_PROCESSES|& & \\
\hline
\hline
\multirow{3}{*}{Contact} & 7 & \multirow{4}{*}{Information theft}  & \lstinline|READ_CONTACTS| & 8,9 & Intents\\
\cline{2-2}\cline{4-6}
\multirow{3}{*}{Extractor}& 8 & & \lstinline|WRITE_EXTERNAL_STORAGE| & 7,9 & Ext. Storage \\
\cline{2-2}\cline{4-6}
& \multirow{2}{*}{9} &  & \lstinline|INTERNET| & \multirow{2}{*}{7,8} & Intents \\
&  &  & \lstinline|READ_EXTERNAL_STORAGE|  &  & Ext. Storage \\
\hline
\hline
 \multirow{2}{*}{Non-colluding} & 10 & \lstinline|-| & - & - & - \\
\cline{2-6}
& 11 & \lstinline|-| & \lstinline|INTERNET| & - & - \\
\hline
%\multirow{7}{*}{Aggregated permissions} & \lstinline|WRITE_EXTERNAL_STORAGE| & \lstinline|READ_CONTACTS| & \lstinline|READ_CONTACTS| \\
% & \lstinline|INTERNET| & \lstinline|READ_SMS| & \lstinline|INTERNET| \\
% &  & \lstinline|SEND_SMS| & \\
% &  & \lstinline|GET_TASKS| & \\
% &  & \lstinline|KILL_BACKGROUND_PROCESSES| & \\
% &  & \lstinline|INTERNET| & \\
%\hline
% & & & \\
%\hline
\end{tabular}
\end{table*}

\subsection{Validation Run}
%TODO A6(Jorge)- Remove mentions to false positives and state the there is no ground truth. We are analysing an unknown problem (with real world apps). Our effort is not to detect and compare against a ground truth, but to extract knowledge from a set of apps we are analysing that might or might not be colluding. Say that there is no state of the art that defines a ground truth of colluding apps we can test againts.
%3-Automatically analyse the app set in prolog.

Table \ref{tab:col_matrix} shows the results obtained from analysing the crafted colluding app set. ``Dark red club'' entries show when we detect collusion potential. As an example, the entry in row 1, column 2 means: the program detects that $app_1$ sends information to $app_2$, and these two apps collude to perform ``information theft''. As we take communication direction into consideration, the resulting matrix is non-symmetric, e.g., there is no entry in row 2, column 1. Additionally, our approach is able to identify transitive collusion attacks (i.e. $app_7$ colluding with $app_9$ through $app_8$).

\begin{table}[ht]
\centering
\caption{Collusion Matrix of the Prolog program. \textcolor{truepositive}{$\clubsuit$} = Information theft. \textcolor{truepositive}{\textdollar} = Money theft. \textcolor{truepositive}{$\spadesuit$} = Service misuse. \textcolor{falsepositive}{$\clubsuit$}, \textcolor{falsepositive}{\textdollar}, \textcolor{falsepositive}{$\spadesuit$} = Benign showing collusion potential.}
\label{tab:col_matrix}
\setlength\tabcolsep{5pt}
\begin{tabular}{| c | c | c |c | c|c | c | c | c | c | c | c |}
 \hline
\textbf{app} & \textbf{1} & \textbf{2} & \textbf{3} & \textbf{4} & \textbf{5} & \textbf{6} & \textbf{7} & \textbf{8} & \textbf{9} & \textbf{10} & \textbf{11}  \\
 \hline
\textbf{1} &  & \textcolor{truepositive}{$\clubsuit$} & & & & & & & & \textcolor{falsepositive}{$\clubsuit$} & \textcolor{falsepositive}{$\clubsuit$}  \\
  \hline
\textbf{2} & &  & & & & & & & & &   \\
  \hline
\textbf{3} & & &  & \textcolor{truepositive}{\textdollar}\textcolor{falsepositive}{$\clubsuit$} & \textcolor{truepositive}{$\spadesuit$} & \textcolor{truepositive}{$\spadesuit$} & & & & &   \\
  \hline
\textbf{4} & & & &  & & & & & & &  \\
  \hline
\textbf{5}  & & & \textcolor{truepositive}{$\clubsuit$} &  &  & & & & & \textcolor{falsepositive}{$\clubsuit$} & \textcolor{falsepositive}{$\clubsuit$}   \\
  \hline
\textbf{6}  & & & \textcolor{truepositive}{$\clubsuit$} & \textcolor{falsepositive}{$\clubsuit$} &  &  & & & & &   \\
  \hline
\textbf{7} & & \textcolor{falsepositive}{$\clubsuit$} & & & & &  & \textcolor{truepositive}{$\clubsuit$} & \textcolor{truepositive}{$\clubsuit$} & \textcolor{falsepositive}{$\clubsuit$} & \textcolor{falsepositive}{$\clubsuit$}  \\
  \hline
\textbf{8} & & & & & & & &  & \textcolor{truepositive}{$\clubsuit$} & &  \\
  \hline
\textbf{9} & & & & & & & & & & &   \\
\hline
\textbf{10} & & & & & & & & & & & \textcolor{falsepositive}{$\clubsuit$}  \\
  \hline
\textbf{11} & & & & & & & & & & &   \\
  \hline
\end{tabular}
\end{table}

`Gold club'' entries show apps flagged as having some collusion potential by our approach but not colluding in reality. For instance, the entry in row 1, column 10 means: the program flags collusion of type ``information theft'' though the set \{$app_1$, $app_{10}$\} is clean. However, they are just exchanging information. As stated in our definition of collusion potential, some benign apps can share access to sensitive resources (e.g. a location being shared from a maps app to a social media app). As we consider all Android available channels as suitable for collusion, in our first approach apps using common channels such as intents with \lstinline|VIEW| or \lstinline|SEND| actions that are very commonly used in Android are also considered to have collusion potential. However, it is unlikely to see apps using these channels for collusion as other apps could have registered to receive the same information. We have taken this fact into consideration when scaling up our methodology (Section \ref{sec:s-u}). %Second, we identify apps that are communicating by sharing access to sensitive resources, but we do not look at how that access is shared. %It must be noted, that the main aim of our method is to point out app sets with collusion potential to reduce the search space and ease the security analyst tasks. 

Overall, our approach identifies all colluding app sets but it also flags eight cases with collusion potential where apps are just collaborating.% Note, that there are no false negatives thanks to the use of a tailored test set: apps only utilise communication methods that our approach is able to identify.%, i.e., our test set is biased.

\section{Scaling up}
\label{sec:s-u}

Our initial methodology takes an app set and finds all potentially colluding app sets. This works as long as the app set to analyze is a  reasonable size (i.e.\ the number of apps that are regularly installed on a regular smartphone, about 20 to 30). However, if the number of applications to analyze is larger, scalability problems arise.

\subsection{Collusion Potential and Computational Complexity}
\begin{figure}[ht]
\begin{center}
\includegraphics[width=9cm]{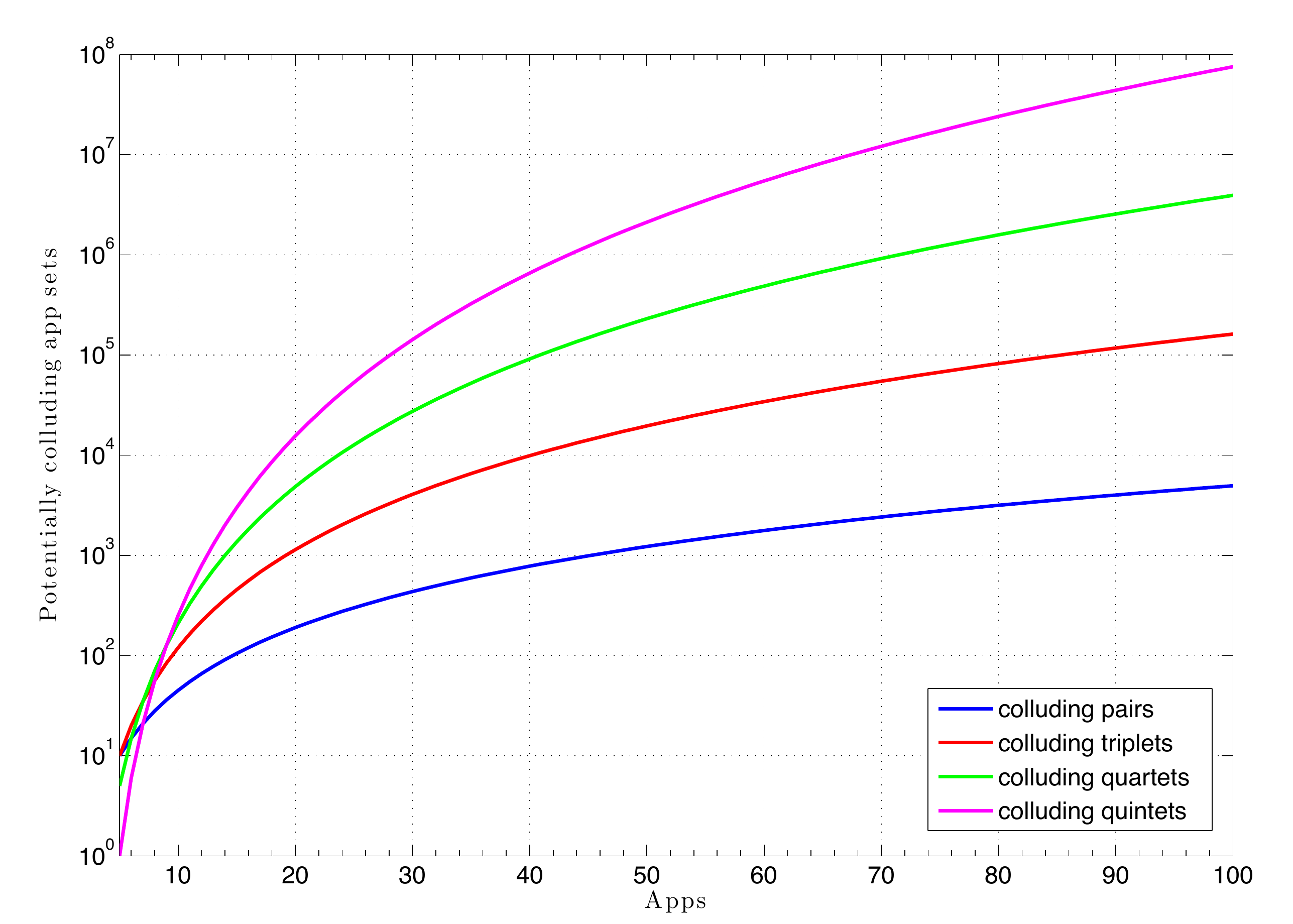}
\caption{Maximum number of potentially colluding app sets that can be found depending on the size of the number of apps analyzed.}
\label{fig:estimation}
\end{center}
\end{figure}

Figure \ref{fig:estimation} shows an estimation of the maximum number of potentially colluding sets depending on the size of the app set to be analyzed. These estimates correspond to the possible combinations of $k$ apps in a group of $n$ apps.%, not taking into account the collusion direction. 

As the number of apps increases, the number of potentially colluding sets increases exponentially. 
%For example, if each application in a group of, say, 100 apps is able to communicate with the other apps in the set, this could produce about one million potential colluding app sets. 
Therefore, if the analyzed apps show a high degree of communication capabilities, the number of potentially colluding app sets will become unmanageably large. 
%for the security analyst and for the system where the analysis is being executed.

\subsection{Managing Complexity}

To address scalability, we improved the Prolog-based methodology considering the way Android works.
Android's design philosophy strongly promotes the use of intents and other IPC communications in order to improve user experience. Consequently, many communication paths detected by our method will be benign, generating alarms for app sets that have collusion potential, but are just collaborating. If we were able to identify and remove these benign communications, then the corresponding potentially colluding sets would be reduced drastically. 

We analyzed communication signatures generated by more than 50,000 apps included in our experiment dataset (Section \ref{sec:experiments}). Up to 40\% of the analyzed apps have the capability to read from and write to external storage. Our approach does not identify specific files accessed in the external storage. Consequently, our crude initial approach would consider all these apps to be  capable of communicating with each other. However, this is not true and fails to represent their real behaviour. We decided to leave out external storage as a communication channel when scaling up our approach. Identification of specific files opened by each app is left for future work. 

Similarly, we filtered out some common intents used by apps to exchange information. Specifically, we have removed the following intent-based communications\footnote{The full list of such intents can be found in our github repository}:
\begin{itemize}
\item Intents that can only be generated by the operating system. These can be found in the Android Open Source Project Git page\footnote{\url{https://android.googlesource.com}}. We have identified 253 intent actions in this category.
\item Intent actions that are created by common and trusted third party applications such as Facebook, Dropbox, etc. These are sent by applications that want to interact with these apps, but only the apps from the same developer (Facebook, Dropbox, etc.) receive them. We can detect them by inspecting intents sent and received by the \textit{clean} apps of our data set -- c.f.\ Section \ref{ssec:dataset} for details.  Intents exhibiting this behaviour will be received by one (e.g. Facebook) or a small number of apps (e.g. apps implementing the Facebook API). To rule out such intents we measured the amount of apps that send the intent divided by the ones that are able to receive it:
\begin{equation*}
c_{intent} = \frac{apps\_sending_{intent} }{apps\_receiving_{intent} }
\end{equation*}
Any intent action included in the aforementioned apps or with a $C_{intent} \geq 5$ has been included in this list. We have obtained 693 intent actions to filter by using this approach.
\item Intents that are used to execute common tasks such as view a document (\lstinline|android.intent.action.VIEW|); send something (\lstinline|android.intent.action.SEND|); or open an application (\lstinline|android.intent.action.MAIN|) are widely used in the Android ecosystem. Some of these are defined in the Android documentation\footnote{\url{http://developer.android.com/reference/android/content/Intent.html}}. These kinds of intents are widely sent and received by apps. As they are declared by many apps, in most cases, the user will be asked to select the app to handle the intent, making the collusion attack infeasible. We have identified 208 intent actions matching these characteristics.
\end{itemize}

%TOTO Probably change the name of this section to something more like. Analysing real world apps.
\section{Experiments}
\label{sec:experiments}
%TODO A6(Jorge)- Remove mentions to false positives and state the there is no ground truth. We are analysing an unknown problem (with real world apps). Our effort is not to detect and compare against a ground truth, but to extract knowledge from a set of apps we are analysing that might or might not be colluding. Say that there is no state of the art that defines a ground truth of colluding apps we can test againts.
We have used our methodology to look for collusion potential in a set of 50,174 apps provided by Intel Security (McAfee). The goal of this analysis was to shed the light on the way Android apps communicate and to test if our approach can deal with high numbers of apps found in the wild. As our approach focuses on specific, selected communication channels, it might happen that apps not flagged by our approach could be colluding as they use a channel which our analysis does not track. 

\subsection{Dataset Description}\label{ssec:dataset}

The dataset contains 50,174 Android apps collected from February 2012 up to February 2016. These apps have been categorized by Intel Security into 3 app categories: \textit{malicious}, \textit{potentially unwanted}, and \textit{clean}. Potentially unwanted apps are typically related to excessive advertising, mild privacy invasions and other misbehaviours which cannot be classified as outright malicious. Apps that are known to lack any malicious behaviour are labelled as clean. 
% This last group included the most prevalent clean apps seen in devices which run our partner's security product for Android. 
Table \ref{tab:intel_apps} shows a summary of the three groups.% including the time interval in where APKs where first and last seen (the same for all three categories).

\begin{table}[ht]
\centering
\caption{Summary of app sets used in our analysis.}
\label{tab:intel_apps}
\begin{tabular}{|l|r|r|r|}
\hline
& \textbf{Malware} & \textbf{Unwanted} & \textbf{Clean} \\
\hline
\textbf{\# of apps} & 13,805  & 13,991 & 22,378 \\
\hline
\textbf{\# of overall installs} & 3,696,720 & 7,656,755 & 21,205,724,533 \\
\hline
\textbf{Avg size in KB} & 3,007.9 & 7,394.52 & 10,208.3 \\
\hline
\end{tabular}
\end{table}
%TODO maybe we could add average, min and max app size. 

\subsection{Usage of Communication Channels}

\begin{figure}[ht]
\begin{center}
\includegraphics[width=9cm]{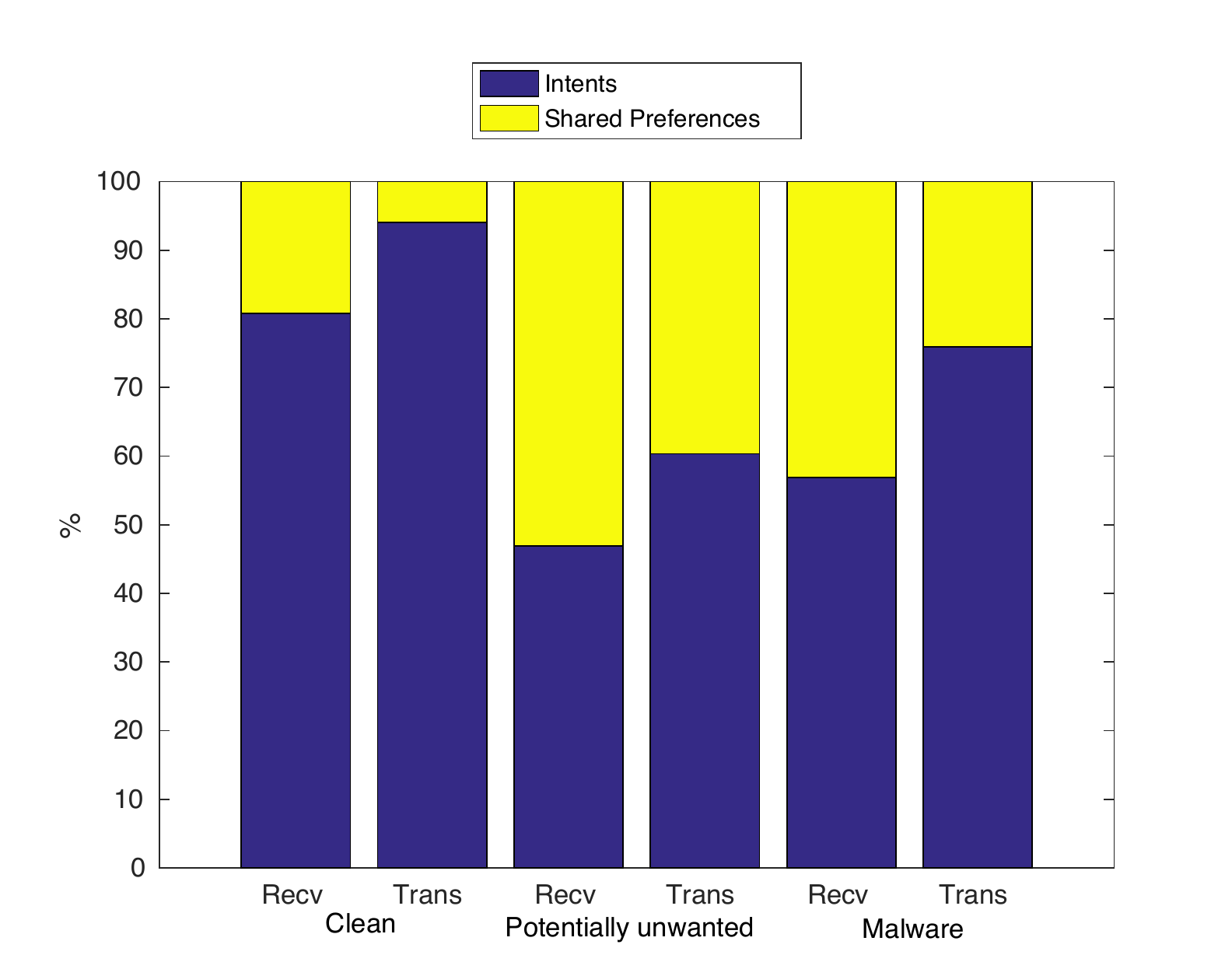}
\caption{Distribution of unique shared preference and intent based communication channels. \emph{Recv} channels are used to receive information. \emph{Trans} channels used to transmit information.}
\label{fig:intentvsshared}
\end{center}
\end{figure}

We first checked if there is a difference between the usage of intents and shared preferences as communication channels. Figure \ref{fig:intentvsshared} shows the distribution of individual channels found in each of the analyzed app sets (after filtering out common intents). Our first observation is that intent based communication is more predominant in the three analyzed app sets. This is expected result because intent based communications is the suggested method for inter-app communication in the Android documentation. 

\subsubsection{Shared Preferences based Channels}

We found a significant difference in the amount of individual channels that use shared preferences for the malicious and unwanted app sets. Shared Preferences are not originally intended for application communication. If a developer wants to make a shared preference file accessible outside of the sandbox, he needs to explicitly override the default flag value (\lstinline|WORLD_READABLE| or \lstinline|WORLD_WRITABLE|). Therefore, it is more likely that the presence of such channels indicates deliberate information sharing rather than mistakes during app development. 

\begin{figure}[ht]
\begin{center}
\includegraphics[width=9cm]{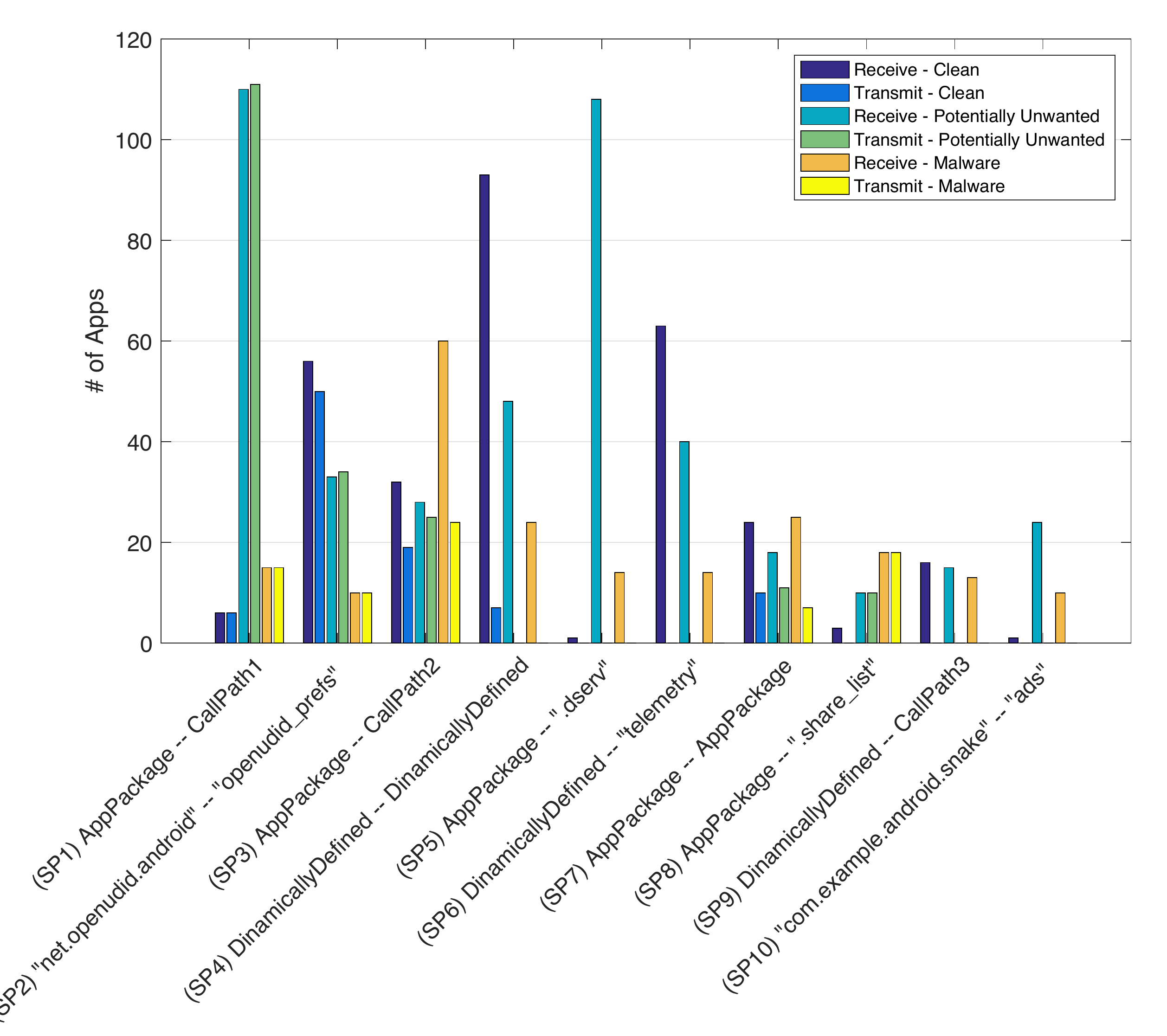}
\caption{Apps using any of the top 10 shared preference channels. AppPackage specifies the application package. CallGraphX describes a call graph that is being used by different apps.}
\label{fig:sharedprefs10}
\end{center}
\end{figure}

We have further analyzed how the most common shared preference channels are used. Figure \ref{fig:sharedprefs10} shows the number of apps in each set using each of the top ten identified channels to send or receive information.

Next we manually analyzed samples of the apps employing these communication channels. We found that apps using shared preferences as channels fall into three categories. First, there are some apps that dynamically define the package and preference file they write to or read from ($SP4$). In this case, each app uses the preference file for a specific purpose and it is not possible to extract a behavioural pattern. Second, some apps access preference files that have the same name as their app package ($SP7$). These apps also exhibit different behaviours so it is not possible to extract a pattern from them. The third category consists of the rest of the channels, which are the ones that can be directly mapped to a string value or a call graph. We have found out that all these channels were related to using specific software libraries. 

The most used shared preference channels in Figure \ref{fig:sharedprefs10} were traced back to five different libraries. Channels $SP1$, $SP5$ and $SP8$ are included inside a SDK that belongs to the Chinese company \emph{Play.cn}. While most apps including it are categorized as malicious or potentially harmful, some of them are considered clean. These apps should be further analyzed to determine their behaviour. Channel $SP2$ is created by the \emph{OpenUDID} library. This library, which is now discontinued, was used to generate a unique identifier that could be shared between different apps. This behaviour puts the user's privacy at risk: it can be used by different apps to correlate if they are installed in the same device. $SP3$ and $SP9$ belong to a library developed by the Chinese company \emph{Baidu}. We have been able to identify a colluding behaviour by apps using this SDK, which is described in detail in Section \ref{sec:moplus}. Channel $SP6$ belongs to apps including the Adobe Air SDK. The preference file is read by a method named \lstinline|getTelemetrySettings|. We did not find any app writing into that file in any of our three app sets. Finally, channel $SP10$ belongs to apps including the \emph{Heyzap} advertising library. Again, no apps in our three datasets were found writing data into that file. 

\subsubsection{Intent based Channels}

\begin{figure}[ht]
\begin{center}
\includegraphics[width=9cm]{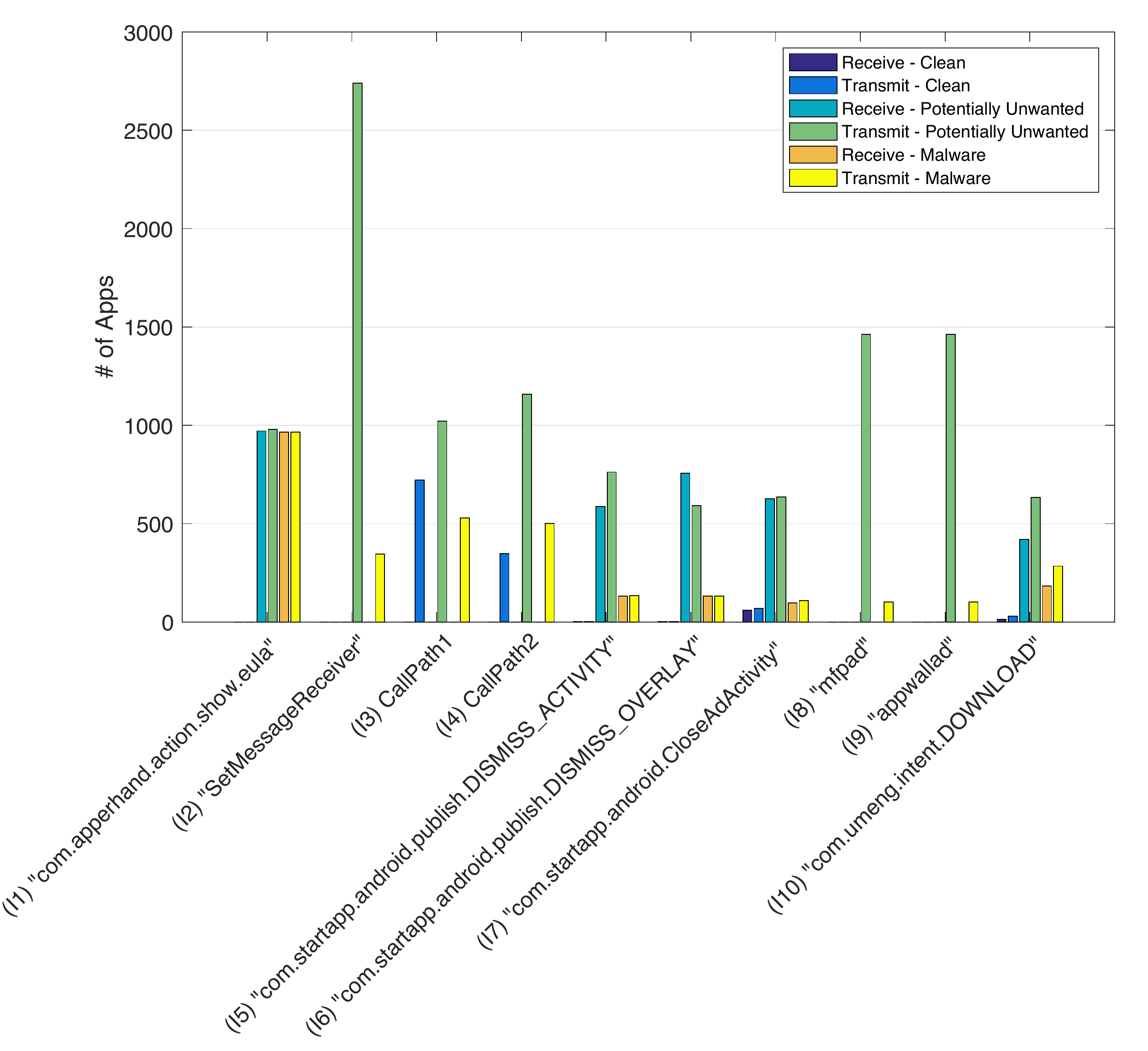}
\caption{Apps using any of the top ten used intent enabled channels. CallGraphX describes a call graph that is being used by different apps. Quoted values are strings.}
\label{fig:intents10}
\end{center}
\end{figure}

Figure \ref{fig:intents10} shows the number of apps that use the most frequent intents in all the three sets of apps. Analyzing intent communication is more challenging than shared preferences communications. Depending on the app component (activity, service or broadcast receiver) used to match an intent, it is not possible to see by static analysis if it is intended for the same app or a different one. Additionally, as seen with $I3$ and $I4$, we have not detected any receiver that uses intent actions defined programatically. This is because components that receive intents are generally defined statically with strings inside the \lstinline|AndroidManifest.xml| file. 

We have found that intent based communications can also be used to help with app classification. In Figure \ref{fig:intents10}, malicious and unwanted apps use inter-app communication channels that are not being used by clean apps. 

As with the shared preferences, we analyzed the origin of the predominant communication channels found in our analysis. Most of them belong to libraries provided by advertisement companies. $I1$ is included in an aggressive ad library known as \emph{apperhand}. $I2$, $I4$, $I8$ and $I9$ belong to the \emph{AirPush} advertisement library. The \emph{Sendroid} ad library includes the communication channel $I3$. Finally, $I5$, $I6$ and $I7$ are included inside the \emph{Startapp} ad library while $I10$ appears in a Chinese library called \emph{umeng}.

\subsection{Collusion Potential}

Using Prolog to analyze collusion potential provides a great deal of flexibility, by simply modifying the Prolog rules to define collusions to look for. We have focused on analyzing collusion potential of app sets formed by 2 or 3 apps that may try to extract the accounts, SMS messages or the device contact list. We have limited the size of app sets to two and three for two reasons. First, it is unlikely that an attacker has the resources to make the user to install more than 3 apps. %If the attacker has this capability then it might also have the necessary resources to deploy more complex attacks, such as using zero-day exploits, etc. 
Second, larger app sets will be composed of smaller subsets. Finding them is just a matter of combining smaller colluding app sets. 
Listing \ref{lst:colpot} shows the Prolog rules required to identify apps with collusion potential that may affect the accounts, SMS messages or the contacts of the device.% The only parameter required is the length of the app sets to be queried. 

\begin{lstlisting}[caption={Selection of collusion potential Prolog rules},basicstyle=\scriptsize,label={lst:colpot}]
coll_accounts(AppA,AppB,Path,Length):- uses(AppA,'GET_ACCOUNTS'), comm(AppA,AppB,Length,_,Path), out_comm(AppB).
coll_contacts(AppA,AppB,Path,Length): uses(AppA,'READ_CONTACTS'), comm(AppA,AppB,Length,_,Path), out_comm(AppB).
coll_sms(AppA,AppB,Path,Length): uses(AppA,'READ_SMS'), comm(AppA,AppB,Length,_,Path), out_comm(AppB).
\end{lstlisting}

\begin{figure}[t]
\centering
\includegraphics[width=9cm]{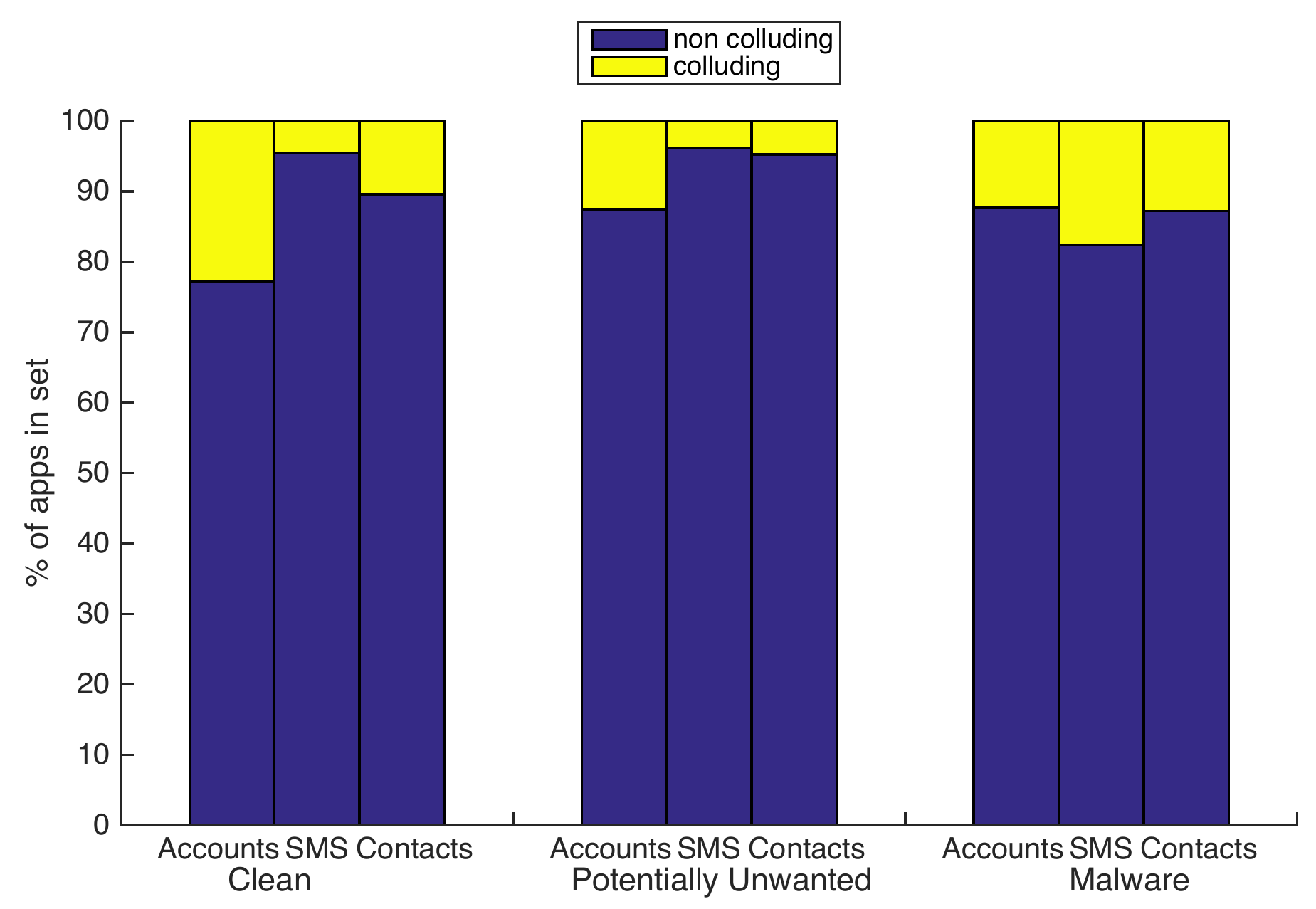}
\caption{\label{fig:percentage}Percentage of applications that access accounts, SMS and contacts with and without collusion potential in each of the datasets.}
\end{figure}

Figure \ref{fig:percentage} shows the percentage of apps inside each set that exhibit collusion potential for any of the analyzed permission-protected resources (accounts, SMS and contacts). Results show that at least the 70\% of apps in each of the datasets do not exhibit collusion potential regarding the analyzed resources. This greatly reduces the number of possible apps to analyze further. Apps inside the malware group exhibit more collusion potential than apps in the other categories (with the exception of accounts in the clean dataset). This is because malware apps generally request more permissions \cite{zhou2012dissecting} and the inclusion of advertisement libraries, as we saw in the previous section.

\begin{figure*}[!ht]
\begin{center}
\includegraphics[width=12cm]{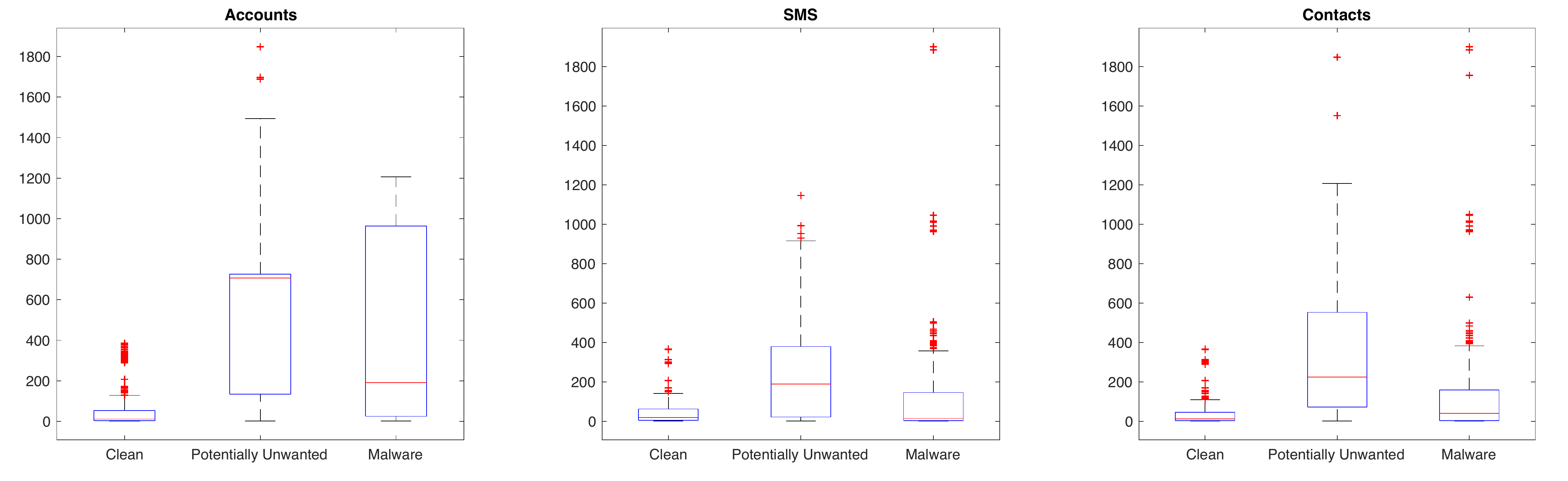}
\caption{Number of potentially colluding app pairs obtained for each app that exhibited collusion potential in each of the sets for each of the analyzed permission-protected resources.}
\label{fig:colpot2}
\end{center}
\end{figure*}

\begin{figure*}[ht]
\begin{center}
\includegraphics[width=12cm]{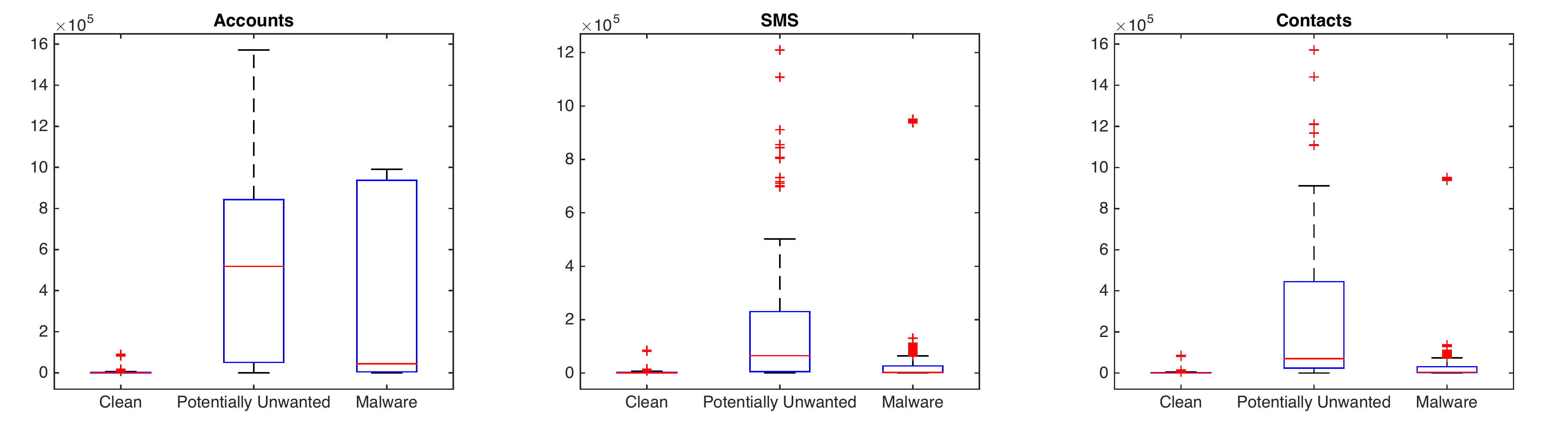}
\caption{Number of the potentially colluding app triplets obtained for each app that exhibited collusion potential in each of the sets for each of the analyzed permission-protected resources.}
\label{fig:colpot3}
\end{center}
\end{figure*}

The potentially unwanted app set includes apps with a less collusion potential. This is a contradictory behaviour. However, when analyzing the amount of apps that have colluding potential inside each of the groups, we found an explanation. Figures \ref{fig:colpot2} and \ref{fig:colpot3} show the number of apps that can receive each sensitive protected resource from an app that has been identified to have collusion potential. Although apps inside the unwanted group have a smaller number of apps capable of leaking sensitive information, they are able to share them with a much higher number of apps than apps in the other categories. This is because apps in this group include aggressive advertisement libraries such as the ones described in the previous section. % A detailed analysis of the apps in showing collusion potential lead us to discovering what we believe is the first case of collusion in the wild. This is explained in more detail in Section \ref{sec:moplus}. 

Apps in the clean set have the smallest number of potentially colluding pairs, while malware apps have a higher number of potentially colluding app pairs regarding accounts. This happens because clean apps request account related permissions more often but communicate with fewer apps, while malware apps require slightly less access to accounts, but communicate with many more apps. 

This pattern remains when analyzing the number of potentially colluding triplets generated by apps with access to the analysed resources. The main difference is the magnitude on the number of colluding triplets, as the number of possible combinations increases.

\subsection{Time Efficiency}

The process required to find app sets with collusion potential is split in two phases: extracting the ASR signatures and executing the Prolog program. The first phase needs to be executed only once per app, as signatures can be stored in a database. The second phase is executed every time the fact database is updated (i.e. when a new app is analyzed). It should be noted that our analysis is not bidirectional as we identify the direction of the information flow. 

\begin{figure}[ht]
\centering
\includegraphics[width=9cm]{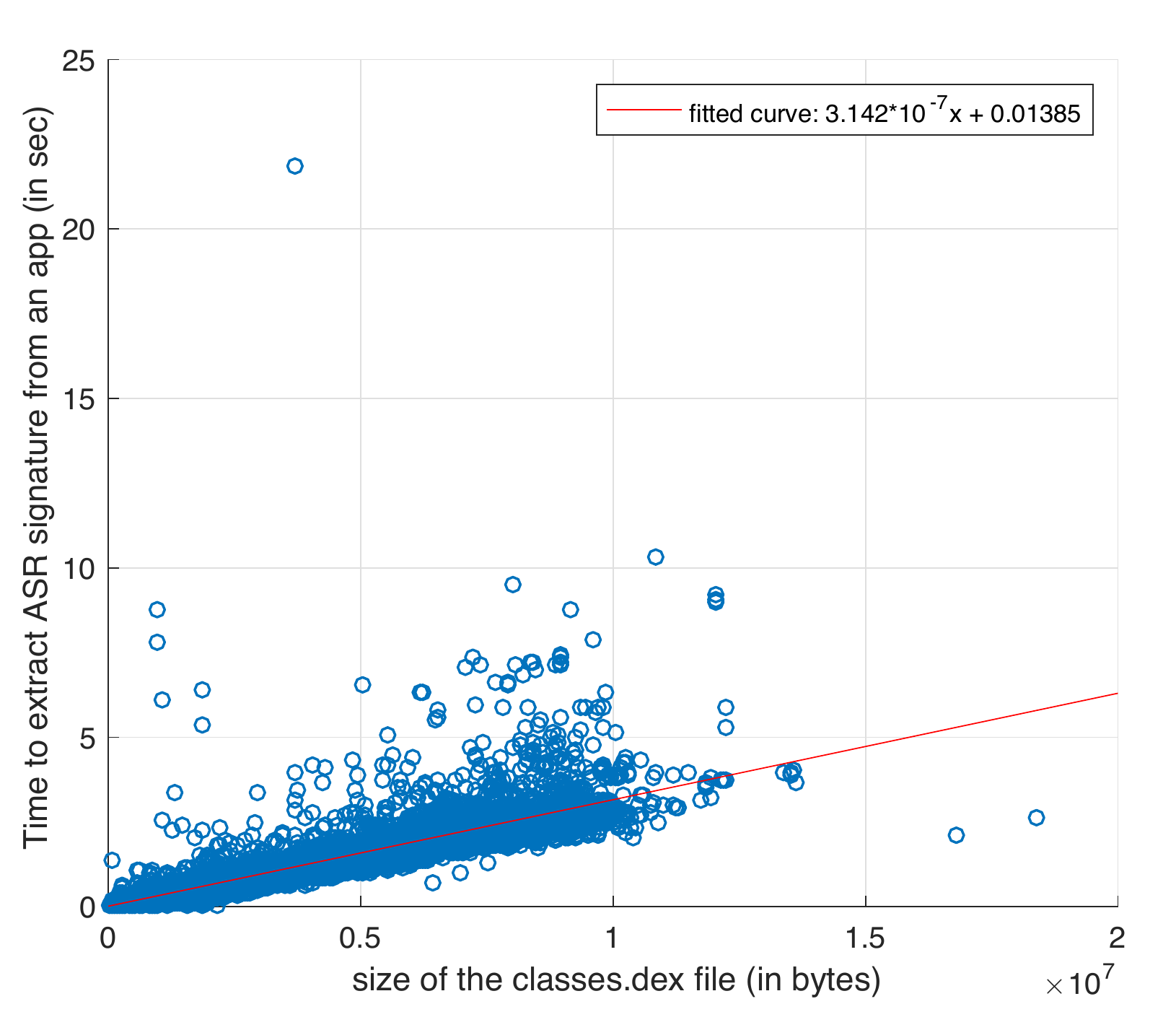}
\caption{\label{fig:fact_time} Time required to extract ASR signatures from an app, depending on the size of the \emph{classes.dex} file.}
\end{figure}
%Goodness of fit:
%  SSE: 2590
%  R-square: 0.7612
%  Adjusted R-square: 0.7612
%  RMSE: 0.5337

Figure \ref{fig:fact_time} shows the time required to extract the ASR signatures from an app, depending on the size of the \emph{classes.dex} file. All experiments were executed on a commodity PC with an Intel Core i5 2.7 GHz processor and 8GB DDR3 RAM. As expected, time grows with the amount of code to be analyzed. Obtained times fit with a linear function with an R-Square of 76\%. For example, a 9.5 Mbyte file requires around 4 seconds of processing. Note that we have not put an emphasis on optimizing the ASR extraction code. 

\begin{figure}[ht]
\centering
\includegraphics[width=9cm]{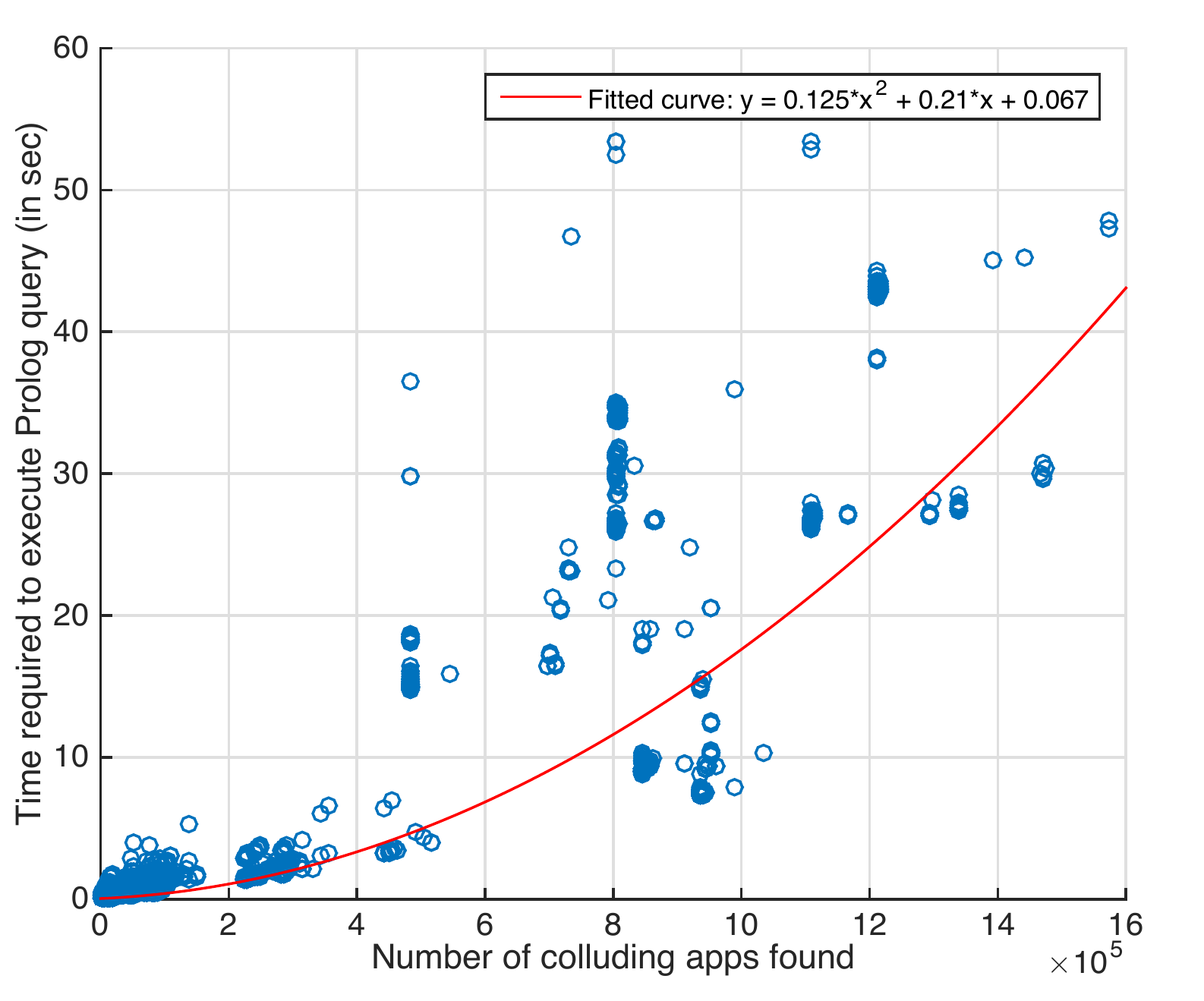}
\caption{\label{fig:prologtime}Time required to execute a Prolog query depending on the amount of potentially colluding app sets the app belongs to.}
\end{figure}

Figure \ref{fig:prologtime} plots the time required to execute a Prolog query depending on the number of colluding sets found for each app. Queries for apps that do not exhibit any colluding behaviour take 30 ms on average. When looking for potentially colluding app pairs, the maximum query time obtained during our experiments was 216 ms. The higher times shown in Figure \ref{fig:prologtime} were obtained when looking for colluding app triplets. Obtained times fit with a polynomial of grade 2 with a R-square of 71\%. Analysis time could be reduced by stopping queries at the first match. In this way, only apps with one match would be analyzed further. 

%Goodness of fit:
%  SSE: 1.524e+11
%  R-square: 0.7108
%  Adjusted R-square: 0.7108
%  RMSE: 975.5

%****************************************
\section{Colluding Behaviour of MoPlus SDK}
\label{sec:moplus}
%****************************************
%TODO (Jorge) Specify more explicitly what makes the Moplus SDK malicious.

During our experiments querying for potentially colluding app pairs, we identified a group of apps that was communicating using both intents and shared preference files. A manual review of the flagged apps revealed that they were sharing information through shared preferences files to synchronize the execution of a potentially harmful payload. This payload was embedded into all the apps through a library, the MoPlus SDK. This library has been known to be malicious since November 2015 \cite{TrendMicro}. However, the collusion behaviour of the SDK was unknown. In the rest of this section, we briefly describe the malicious behaviour of the SDK and provide a more detailed analysis of its colluding behaviour. 
To the best of our knowledge, this is the first instance of collusion found in the wild. 

\subsection{Malicious Payload}

The MoPlus SDK has the ability to open a local HTTP server on the user device. This enables the attacker to perform a series of malicious operations including:
\begin{itemize}
\item Send arbitrary intents received via the command and control (C\&C) server.
\item Obtain sensitive information from the user’s device, including the user location and the IMEI.
\item Install apps silently in rooted devices.
\item Add contacts received from the C\&C server.
\end{itemize}

The malicious payload embedded inside the MoPlus SDK inherits all permissions requested by the app. As these are chosen by the app developer, which may differ from the SDK developer, it is possible that an app including the SDK does not have the necessary permissions to execute all the library's malicious payload. The colluding behaviour of the MoPlus SDK aims to avoid this problem by identifying which of the apps that include the MoPlus SDK and are installed in a device have most access to system resources.

\subsection{Colluding Behaviour}
\label{sec:overview}

The detected colluding behaviour differs from the standard colluding behaviour studied in most app collusion research \cite{schlegel2011soundcomber,marforio2012analysis}. In a nutshell, all apps including the MoPlus SDK that are running on a device will talk to each other to check which of the apps has the most privileges. This app will then be chosen to execute the local HTTP server able to receive commands from the C\&C server, maximizing the effects of the malicious payload.

The MoPlus SDK includes the \lstinline|MoPlusService| and the \lstinline|MoPlusReceiver| components. In all analyzed apps, the service is exported. In Android, this is considered to be a dangerous practice, as also other apps will be able to call and access this service. However, in this case it is a feature used by the SDK to enable communication between its apps.

The colluding behaviour is executed when the MoPlusService is created (\lstinline|onCreate| method). This behaviour is triggered by the MoPlus SDK of each app and can be divided in two phases: establishing app priority and executing the malicious payload. In the next sections, we will describe this behaviour in detail with reconstructed code samples. These have been obtained by reconstructing part of the code from the Baidu Searchbox app with MD5 062f91b3b1c900e2bc710166e6510654 signature. Locations of different payloads may differ from app to app, as code is generally obfuscated by using Proguard.

\subsubsection{Establishing app priority}

During SDK initialization, the \lstinline|MoPlusService| is created inside each app with the MoPlus SDK. The service executes three checks (Listing \ref{lst:priority}):
\begin{enumerate}
\item The version of the MoPlus SDK is checked against a value stored in a preference file (lines 3 to 5). %This check does not change the behaviour of the program.
\item The SDK looks for the tag \lstinline|DisableService| inside the AndroidManifest (\lstinline|!a(Context)|, line 8). If it is found, it will not continue to execute.
\item The SDK checks if the app executing the SDK has all the necessary components of the SDK and the minimum permissions required by the SDK have been granted (\lstinline|j(Context)|, line 8). The minimum permissions required to continue execution are: \lstinline|INTERNET|, \lstinline|READ_PHONE_STATE|, \lstinline|ACCESS_NETWORK_STATE|, \lstinline|BROADCAST_STICKY|, \lstinline|WRITE_SETTINGS|, \lstinline|WRITE_EXTERNAL_STORAGE|, \lstinline|SET_ACTIVITY_WATCHER|, \lstinline|GET_TASKS|.
\end{enumerate}

\begin{lstlisting}[caption={Code used to check for execution conditions. This code is included in the class com.baidu.android.moplus.util.a.},label={lst:priority},basicstyle=\scriptsize,firstline=1,numbers=left,xleftmargin=1.0em,framexleftmargin=1.0em]
SharedPreferences localSharedPreferences = paramContext.getSharedPreferences("pst", 0);
int i = c(paramContext, paramContext.getPackageName());
int j = localSharedPreferences.getInt("pr_v", 0);
SharedPreferences.Editor localEditor1;
if ((j < i) || (paramBoolean)){
	Log.d("Utility", "oldVCode=" + j + " vcode=" + i + " isForce " + paramBoolean);
	localEditor1 = paramContext.getSharedPreferences(paramContext.getPackageName() + ".push_sync", 1).edit();
	if ((!a(paramContext)) && (j(paramContext)))
    	break label197;
	localEditor1.putLong("priority", 0L);
}
\end{lstlisting} 

If any of these checks fail, the service assigns itself a zero priority inside a preference file readable by the rest of the apps installed in the system (line 11). The name of the preference file is created adding the extension \lstinline|.push_sync| to the app package name. The SDK uses the \lstinline|WORLD_READABLE| flag to save the file so other apps can access it.

If the three checks hold, the service executes the method \lstinline|f(Context)|. This method computes a priority to the app that depends on several factors (Listing \ref{lst:checks}). 

\begin{lstlisting}[caption={Code used by MoPlus SDK to assign priority execution to each app \lstinline|MoPlusService|. This code is included in the class com.baidu.android.moplus.util.a.},label={lst:checks},basicstyle=\scriptsize,firstline=1,numbers=left,xleftmargin=1.0em,framexleftmargin=1.0em]
public static long f(Context paramContext){
	long l1 = 0L;
	if (paramContext == null)
		return l1;
	if (!g(paramContext, paramContext.getPackageName()))
		l1 += 1L;
	long l2 = l1 << 1;
	if (!i(paramContext))
		l2 += 1L;
	long l3 = l2 << 1;
	if (!f(paramContext, paramContext.getPackageName()))
		l3 += 1L;
	long l4 = l3 << 1;
	if (d(paramContext, paramContext.getPackageName()))
		l4 += 1L;
	long l5 = l4 << 1;
	if (p(paramContext))
		l5 += 1L;
	long l6 = l5 << 1;
	if (b(paramContext, paramContext.getPackageName()))
		l6 += 1L;
	return 0x79000000000000 | (l6 | 0xFF & i(paramContext, "moplus_addon_priority") << 40);
}
\end{lstlisting}

These include, from lowest to highest priority:
\begin{enumerate}
\item Several meta-data values from the manifest (lines 3 to 15): \lstinline|DisableLocalServer|, \lstinline|DisableStatistic|, \lstinline|DisableApplist|, \lstinline|isBaiduApp|.
\item Access to the contact lists (lines 17 and 18).
\item If the app is part of the system image (l. 20 and 21).
\item A priority value included in the manifest (line 22).
\end{enumerate}

\begin{figure}
\centering
\includegraphics[width=9cm]{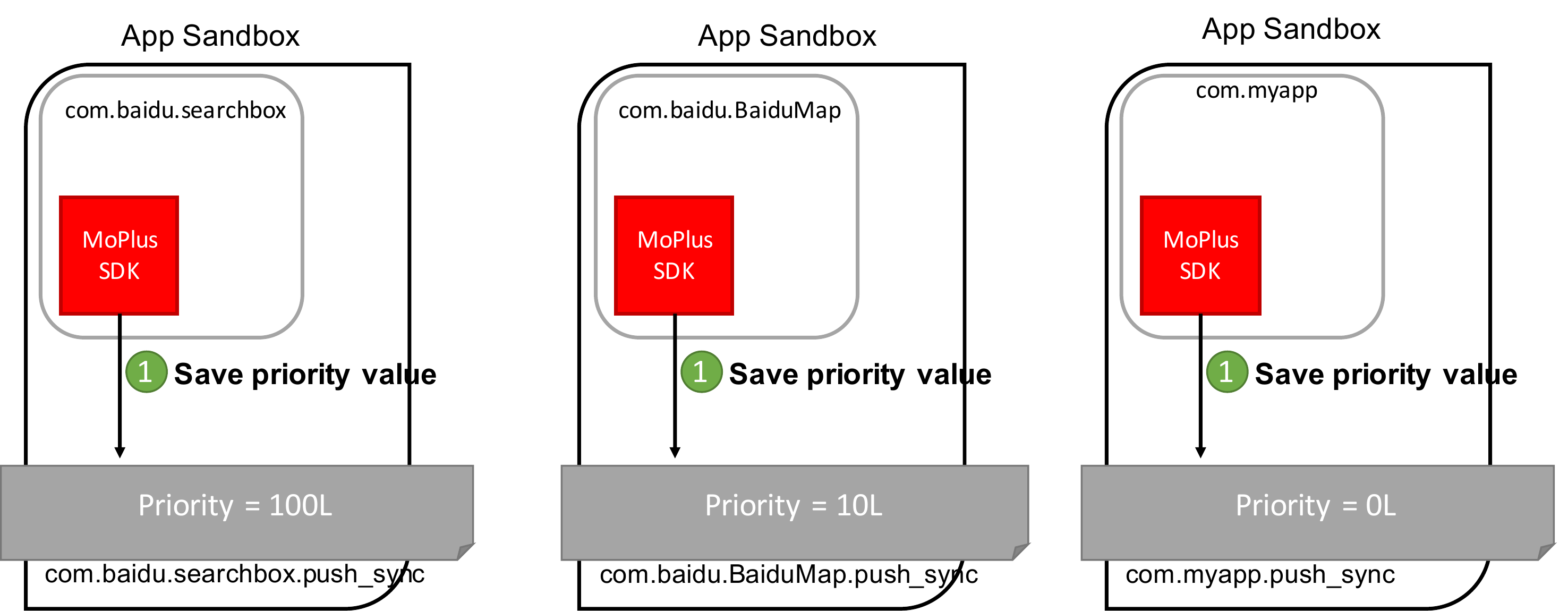}
\caption{\label{fig:fig1}Phase 1 of the colluding behaviour execution. Each app saves a priority value that depends on the amount of access it has to the system resources. Priority values are shown for the sake of explanation.}
\end{figure}

The obtained priority is saved in the preference file with \lstinline|push_sync| extension. This behaviour is executed by all apps including the MoPlus SDK (Figure \ref{fig:fig1}). 
%In this way, each app will hold a shared preference file storing the app's priority (Figure \ref{fig:fig1}).

\subsubsection{Executing the malicious payload}

After the priority has been obtained and stored, the service method \lstinline|OnCreate()| calls the method \lstinline|a(Context,long)| (Listing \ref{lst:intent}) to create and broadcast a new intent object.

\begin{lstlisting}[caption={Creation of a new intent object. The method \lstinline|a| is used to broadcast it. This code is included in the class com.baidu.android.moplus.util.a.},label={lst:intent},basicstyle=\scriptsize,firstline=1,numbers=left,xleftmargin=1.0em,framexleftmargin=1.0em]
public static void a(Context paramContext, long paramLong){
	Context localContext = paramContext.getApplicationContext();
	Intent localIntent = c(localContext);
	localIntent.setPackage(d(localContext));
	a(localContext, localIntent, paramLong);
}
\end{lstlisting}

The \lstinline|localIntent| value is obtained from the execution of the method \lstinline|c(Context)| (line 3). This method creates the intent that will start the \lstinline|MoPlusReceiver| (Listing \ref{lst:action}).

\begin{lstlisting}[caption={Creation of a new intent object. The method \lstinline|a| is used to broadcast it. This code is included in the class com.baidu.android.moplus.util.a.},label={lst:action},basicstyle=\scriptsize,firstline=1,numbers=left,xleftmargin=1.0em,framexleftmargin=1.0em]
 public static Intent c(Context paramContext){
	Intent localIntent = new Intent("com.baidu.android.moplus.action.START");
	localIntent.addFlags(32);
	localIntent.putExtra("method_version", "V1");
	return localIntent;
}
\end{lstlisting}
 
The call to \lstinline|d(Context)| (line 4) looks for the app package with highest priority through the method \lstinline|a(Context,String,String)| (Listing \ref{lst:forward}).

\begin{lstlisting}[caption={Method that returns the app package with highest priority. This code is included in the class com.baidu.android.moplus.util.a.},label={lst:forward},basicstyle=\scriptsize,firstline=1,numbers=left,xleftmargin=1.0em,framexleftmargin=1.0em]
public static String d(Context paramContext){
	return a(paramContext, ".push_sync", "priority");
}
\end{lstlisting}

The method \lstinline|a(Context,String,String)|looks for all the packages that are able to answer the Intents included in the MoPlus SDK (Listing \ref{lst:search}, lines 3 to 6): \lstinline|com.baidu.android.moplus.action.START| and \lstinline|com.baidu.android.pushservice.action.BIND_SYNC|.

\begin{lstlisting}[caption={Method that inspects all shared preference files of packages that answer the MoPlus SDK actions. This code is included in the class com.baidu.android.moplus.util.a.},label={lst:search},basicstyle=\scriptsize,firstline=1,numbers=left,xleftmargin=1.0em,framexleftmargin=1.0em]
public static String a(Context paramContext, String paramString1, String paramString2){
  List localList = h(paramContext);
  if ((localList == null) || (localList.size() <= 1)){
    localObject1 = paramContext.getPackageName();
    return localObject1;
  }
  long l1 = paramContext.getSharedPreferences(paramContext.getPackageName() + ".push_sync", 1).getLong("priority", 0L);
  String str = paramContext.getPackageName();
  Iterator localIterator = localList.iterator();
  while (localIterator.hasNext()){
    localObject2 = ((ResolveInfo)localIterator.next()).activityInfo.packageName;
     SharedPreferences localSharedPreferences2 = paramContext.createPackageContext((String)localObject2, 2).getSharedPreferences((String)localObject2 + paramString1, 1);
    ...
  }
}
\end{lstlisting}

For each package found, it inspects the contents of the \lstinline|push_sync| file to get its priority, returning the package name of the one with highest priority (Figure \ref{lst:search}, line 10 to end). The intent to be launched is configured so only receivers in the returned package can receive it (Listing \ref{lst:intent}, line 4). Finally, the method a(Context, Intent, long) (Listing \ref{lst:launch}) cancels previous intents being registered (to avoid launching the service more than once) and sends the intent after a delay passed as a parameter.

\begin{lstlisting}[caption={Method that cancels previous intents matching the service, and registers a new intent to be launched.},label={lst:launch},basicstyle=\scriptsize,firstline=1,numbers=left,xleftmargin=1.0em,framexleftmargin=1.0em]
public static void a(Context paramContext, Intent paramIntent, long paramLong){
	PendingIntent localPendingIntent = PendingIntent.getBroadcast(paramContext, 0, paramIntent, 268435456);
	AlarmManager localAlarmManager = (AlarmManager)paramContext.getSystemService("alarm");
	localAlarmManager.cancel(localPendingIntent);
	localAlarmManager.set(3, paramLong + SystemClock.elapsedRealtime(), localPendingIntent);
}
\end{lstlisting}

\begin{figure}[ht]
\centering
\includegraphics[width=9cm]{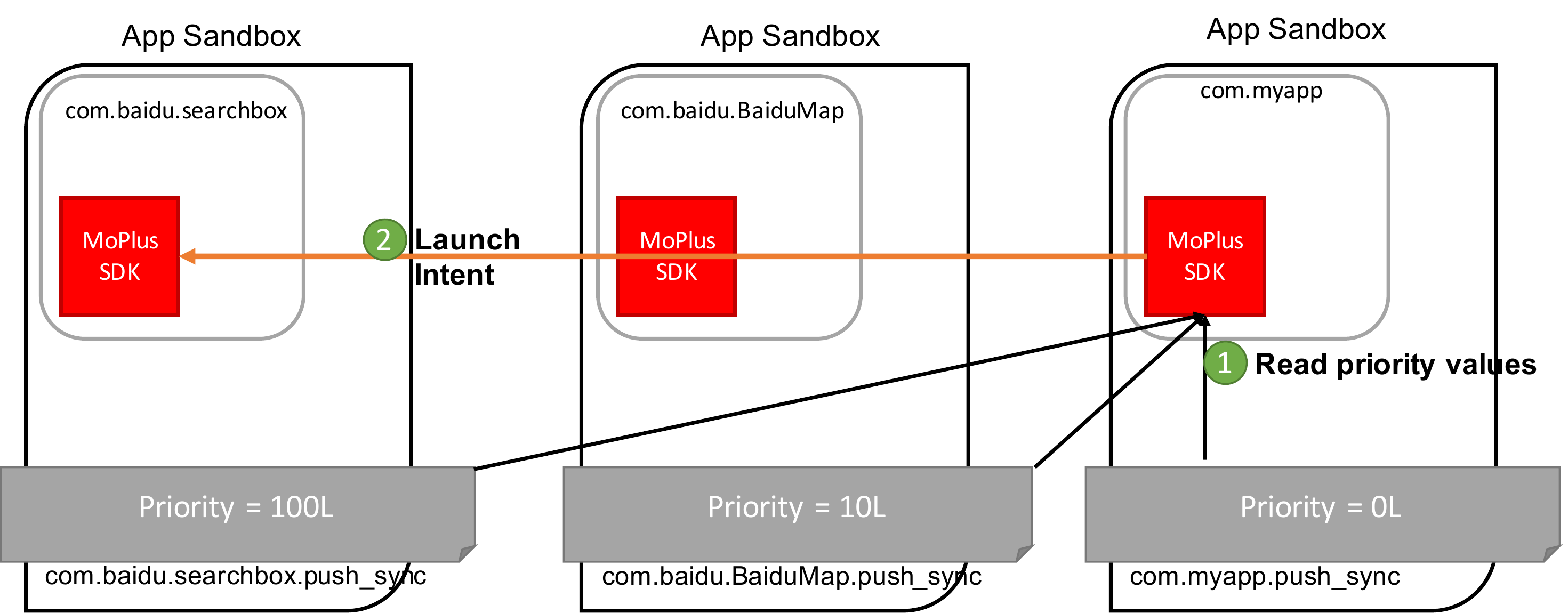}
\caption{\label{fig:fig2} Phase 2 of the colluding behaviour execution. Each app checks the \lstinline|WORLD_READABLE| \lstinline|SharedPreference| files and sends and intent to the app with highest priority}
\end{figure}

The described behaviour is executed by all apps that include the MoPlus SDK libraries (Figure \ref{fig:fig2}).

%****************************************
%     RELATED WORK
%****************************************
\section{Related Work}
\label{sec:related}

Countermeasures for fighting collusion attacks can be grouped into two categories: static analysis and Android OS extensions. %To the best of the author's knowledge dynamic analysis approaches have not been considered yet for this particular problem in Android. 
 ComDroid \cite{chin2011analyzing} is a static analysis tool that looks for confused deputies through \emph{Intents}. {\K}droid \cite{avocs16} detects collusion via software model-checking a set of Android apps utilising the \K \, framework. PermissionFlow uses taint analysis to automatically detect inter-application permission leaks \cite{sbirlea2013automatic}. In their work they found that more than 50\% of the top 313 apps (in 2012) actively used inter-component information flows and four of them leaked permissions to other apps. Our work differs from PermissionFlow in our lack of taint analysis and our consideration of channels that may be used specifically for collusion (e.g. shared preferences). Taint analysis allows PermissionFlow to be more precise, but at the same time more computationally costly. Our system could be used to filter out app sets without colluding potentially, focusing the more computationally complex analysis on those that exhibit collusion potential.
 
In contrast to these,
XManDroid \cite{bugiel2011xmandroid}, TrustDroid \cite{bugiel2011practical} and \cite{jing2016checking} extend the Android OS by providing finer control over app communications. These extensions identify possible communication paths between apps and allow to define policies that control how they exchange information. These are similar to the ones provided by the \emph{Intent Firewall} included as a component, not enabled, of recent Android versions \cite{Yagemann2016}. TrustDroid provides additional controls to monitor the file system and network connections. However, none of them provide a monitoring system for shared preference files and other covert channels. As we have found during our research, these communication channels are also a viable means of communication between colluding apps. 

%In \cite{suarezcompartmentation} the authors analyze several compartmentalization strategies to minimize the risk of app collusion. The assumption is that a device can isolate app groups by forbidding any communication between apps in different groups. Their results show that two or three app compartments greatly reduce the risk posed by a set of 20 to 50 apps. %In order to reduce the risk further, the amount of app compartments must be increased exponentially.

Bartel et al. \cite{apkcombiner2015} propose the tool \textit{APKCombiner} which joins two apps into a single APK file. In this way, a security analyst can use IPC analyzers to analyze the IAC mechanisms. Their evaluation over a set of 3000 apps shows that the approach is valid, as it is capable of joining together 88\% of the possible app pairs. The average time required to join two apps is three minutes. This makes it hard to use for practical large-scale app analysis.% However, it is still a very useful technique to analyse potentially colluding pairs when they are identified by our tool.

\section{Conclusions}
\label{sec:conclusion}

Detecting app collusion on a large scale is a challenging task due to the sheer amount of possible app combinations and communication channels. We have presented a method to analyze large sets of apps to look for collusion potential. Our method is based on a lightweight analysis of apps that extracts \emph{ASR} signatures. These are transformed into Prolog facts, so logic programming can be used to identify collusion potential between apps in an efficient way. 

We have validated our approach against an artificial set of apps and tested it against a large dataset of 'in the wild' apps. 
Our results show that malicious apps use inter-app communications in a different way than clean ones. Malware classification methods could take advantage of this fact to increase their accuracy and to detect collusion. 

A manual analysis of some of the apps flagged by our detection system allowed us to identify the first known case of collusion in the wild. This discovery demonstrated the risk of using untrusted or maliciously modified SDKs. The designers of these app communication scheme even considered the possibility of the SDK being included as part of a system image. Identified colluding apps synchronize with each other (by sharing a priority value) activating only the service within the more privileged app. This finding demonstrates the need to focus on alternative methods of communication when looking for colluding apps. Although intent based communications are more common, other means of communication (such as the shared preferences and covert channels) should also be considered. Our future direction of work aims to explore these communication channels and to use formal verification methods to automatically analyze apps flagged by our system. 

\section*{Acknowledgment}

This work has been supported by UK Engineering and Physical Sciences Research Council (EPSRC) grant EP/L022699/1.

\bibliographystyle{IEEEtran}
\bibliography{biblio}
\vspace{-1.5cm}
\end{document}